\documentclass[journal]{IEEEtran}
\usepackage{blindtext}
\usepackage{graphicx}
\usepackage{amsmath,amsfonts}
\usepackage{amsthm}
\newtheorem{definition}{Definition}[section]

\usepackage{url}
\usepackage{algorithmicx}
\usepackage[ruled]{algorithm}
\usepackage{algpseudocode}
\usepackage{algpascal}
\usepackage{algc}
\usepackage{algorithm}
\usepackage{caption}
\usepackage{float}
\usepackage[caption = false]{subfig}
\usepackage[font=footnotesize,labelfont=bf]{caption}
\usepackage{color}
\usepackage{cases}
\usepackage{amssymb}
\usepackage[table]{xcolor}

\newcommand{\F}{\mathbb{F}}

\ifCLASSINFOpdf
\else
\fi

\begin{document}

\title{A Compact 16-bit S-box over Tower Field $\F_{(((2^2)^2)^2)^2}$ with High Security}

\author{Bahram~Rashidi and Behrooz Khadem{}
	\thanks{Bahram~Rashidi is with Department of Electrical Engineering, Faculty of Engineering, Ayatollah Boroujerdi University, Boroujerd, 69199-69411, Iran e-mail: b.rashidi@abru.ac.ir and Behrooz Khadem is with Faculty of Computer Engineering, communication and network security, Imam Hossein Comprehensive University, Tehran, Iran email: bkhadem@ihu.ac.ir}}

\maketitle

\begin{abstract}
\boldmath
This paper introduces a compact and secure 16-bit substitution box (S-box) designed over the composite field $\F_{(((2^2)^2)^2)^2}$, optimized for both hardware efficiency and cryptographic robustness. The proposed S-box decomposes operations into subfields, leveraging a tower field architecture. This enables significant hardware reduction through optimized field inversion and a low-cost affine transformation. Security evaluations confirm resilience against linear, differential, algebraic and DPA attacks, validated via metrics including Nonlinearity (32512), Differential Uniformity (4), Algebraic Degree (15), Transparency order (15.9875) and SNR (0.34e-08). The hardware results, in 65 nm CMOS technology, show the proposed 16-bit S-box has lower hardware resources consumption and lower critical path delay (CPD) than those of other 16-bit S-boxes. By integrating high algebraic complexity with resource-efficient structures, this work addresses the growing demand for scalable cryptographic primitives in data-sensitive applications, demonstrating that larger S-boxes can enhance security without proportional hardware costs. The results underscore the viability of composite field-based architectures in balancing security and efficiency for modern block ciphers.

\end{abstract}


\IEEEpeerreviewmaketitle

\section{Introduction}\label{sec intro}

The transfer of digital data is growing rapidly. At the same time, the need for security of this digital data is becoming increasingly critical. A significant amount of sensitive information is now stored and processed digitally, making it essential to implement effective security protocols and necessary security points must be considered to prevent attacks and data breaches. Encryption by block ciphers is a method that is commonly used to increase data security, so that the original information cannot be obtained without knowing the secret key used by the encryption function. Substitution boxes (S-boxes) are one of the main components of block ciphers (particularly in structures like the Substitution-Permutation Network (SPN) and Feistel networks) that play a vital role in adding an additional layer of security by introducing nonlinearity and complexity to the encryption process. S-boxes are the most critical nonlinear component in many block ciphers and hash functions. Their primary function is to mix input bits, making them indispensable for ensuring the cipher’s security. Designers must specifically ensure that their S-boxes are resistant to linear, differential, and boomerang attacks, which represent the primary threats to S-box security today \cite{r17}. Larger S-boxes introduce significantly higher levels of nonlinearity and algebraic complexity compared to smaller S-boxes with similar structures, such as those used in the AES S-box. This enhancement allows for a substantial increase in the computational complexity required to attack the algorithm, without the need to add more rounds. Moreover, the hardware area required for these larger S-boxes only experiences a minimal increase when compared to their 8-bit S-boxes.

In recent years, the S-boxes are designed based on field inversion, look-up table (LUT), heuristic method, and pseudorandom approaches. Various S-box structures have been introduced in the literature, as highlighted in works ranging from \cite{r17} to \cite{r23}. For instance, in \cite{r17} a 10-bit S-box generated from a Feistel construction is proposed. The subpermutations are generated by a 5-cell cellular automaton based on a unique well-chosen rule and bijective affine transformations. A chaotic S-box leveraging the intertwining logistic map and bacterial foraging optimization was developed in \cite{6}. In \cite{11}, a chaotic S-box was designed using a difference distribution table. Meanwhile, \cite{13} proposed an S-box based on a heuristic method inspired by the bee waggle dance. Additionally, \cite{14} introduced an S-box derived from projective linear groups on the projective line and permutation triangle groups. The combination of artificial bee colony optimization and chaotic maps was employed in \cite{16} for S-box design. A simpler S-box structure was implemented in \cite{22} using cubic polynomial mapping. Furthermore, \cite{Dragan} explored the generation of random bijective S-boxes using a one-dimensional discrete chaotic map. In \cite{r24} a 5-bit S-box is designed using a chaotic mapping theory to offer a random behavior of the element in the S-box. In works \cite{BRI} and \cite{BRI2} two lightweight 8-bit inversion-based S-boxes are presented. In \cite{r2} a widely applicable method for constructing lightweight S-boxes with differential and linear branch numbers greater than 2, from smaller S-boxes is presented. Using structures such as the Feistel, Lai-Massey, unbalanced-MISTY and unbalanced-Bridge structure. The 16-bit S-boxes are presented in works \cite{r4}, \cite{r5}, \cite{r6}, \cite{r7}, \cite{r9}, \cite{r11}, \cite{r12}, \cite{r19}, and \cite{r23}. For example, in \cite{r9} implement the 16-bit S-box with less circuit area over the composite fields $ \F_{(2^4)^2} $ by the three types of irreducible polynomials. In \cite{r6} an extension of Canright’s tower field algebraic optimizations for in both polynomial and normal bases to computing the 16-bit S-box in $\F_{(((2^2)^2)^2)^2}$ is presented.

The S-box, being a nonlinear element in block ciphers, plays a crucial role in ensuring the security of block ciphers against various attacks. Therefore, it is essential to choose an S-box that offers both high security and area efficiency for the implementation of cryptographic systems. In this paper, we design a compact 16-bit S-box with a high security level. The proposed S-box is based on field inversion in composite field $\F_{(((2^2)^2)^2)^2}$ with a low-cost affine transformation. The proposed structure has acceptable hardware and security properties compared to the other works. The contributions of this paper are summarized as follows:\vspace{0.1cm}

\begin{itemize}
	
	\item {The design and implementation of a low-cost and high security 16-bit S-box is proposed.}\vspace{0.1cm}
	
	\item {The core element used in the design of the proposed S-box is a field inversion unit in the composite field $\F_{(((2^2)^2)^2)^2}$.}\vspace{0.1cm}
	
	\item {The original field $\F_{2^{16}}$ is converted into the tower fields such as $\F_{(((2^2)^2)^2)^2}$, $\F_{((2^2)^2)^2}$, and $\F_{(2^2)^2}$. The operations in these sub-fields are designed to be more compact compared to those in the original field.}\vspace{0.1cm}	
	
	\item {A low-cost affine transformation with low hardware and low delay is used in the S-box.}\vspace{0.1cm}

	\item {The security level of the proposed S-box is acceptable based on security analysis.}\vspace{0.1cm}
	
\end{itemize} 

The rest of the paper is organized as follows. Tower construction with normal bases is presented in Section \ref{TC}. Section \ref{CF} presents the composite field arithmetic and the proposed structures. Proposed structure of 16-bit S-box is described in Section \ref{PS}. Sections \ref{S} and \ref{H} present the security and hardware results. Finally, the paper is concluded in section \ref{sec-end}.

\section{Tower Construction with Normal Bases}
\label{TC}

To achieve efficient implementation, it is natural to consider using a normal basis of $\F_{2^{16}}$ over $\F_{2}$. However, while direct implementation of inversions in $\F_{2^{16}}$ using the normal basis is possible, it often results in high hardware and time complexities. To mitigate these challenges, we adopt the isomorphic tower construction of $\F_{2^{16}}$, which offers improved performance. A simple approach is the construction of the finite field $\F_{2^{16}}$ as an algebraic extension of the prime field $\F_{2}$ , using the primitive polynomial $ p(x) = x^{16} + x^5 + x^3 + x^2 + 1 $. The root of $ p(x) $ is a primitive element of $\F_{2^{16}}$, and will be denoted with $ \omega $, i.e. $ p(\omega) = 0 $. Hence, we write the polynomial basis of $\F_{2^{16}}$ over $\F_{2}$ as $ P = {1, \omega, \omega^2,... , \omega^{15}} $. To find a normal basis we must first find a normal element; Let $ \theta $ be an element in $\F_{q^m}$. Then $ \theta $ is a normal element of $\F_{q^m}$ / $\F_{q}$ if and only if the polynomials $ x^m-1 $ and $\sum_{i=0}^{m-1} \theta^{q^i}x^i$ in $\F_{q^m}[x]$ are relatively prime. Once we have found a normal element $\theta \in \F_{2^{16}}$, we can write down the normal basis: $N = {\theta, \theta^2,... \theta^{2^{15}}}$. 

Here, the normal element with lowest complexity is the element $\omega^{1091}$, where $ \omega $ is the root of irreducible polynomial $ p(x) $ \cite{1}. We use this element to generate the normal basis of $\F_{2^{16}}$ over $\F_{2}$:\vspace{0.3cm} 

$N = {\theta, \theta^2,... \theta^{2^{15}}}$, where $\theta=\omega^{1091} \in \F_{2^{16}}$ and $ p(\omega) = 0 $. An element $A \in\F_{2^{16}}$ is now represented as:\vspace{0.1cm}

$A=\sum_{i=0}^{15} a_i\theta^{2^i}~~~a_i\in\F_{2}.$\vspace{0.3cm} 

The tower construction $\F_{(((2^2)^2)^2)^2}$ uses extensions of degree two on each level of the tower, hence we need an irreducible polynomial of degree two on each step. The tower construction for $\F_{(((2^2)^2)^2)^2}$ is summarized in Table \ref{T:T1}.

\begin{table}[!t]
	\centering
	\captionsetup{justification=centering}
	{\scriptsize	
		\caption{{\footnotesize Tower construction of $\F_{(((2^2)^2)^2)^2}$, $ \F_{2}\xrightarrow{e(x)}\F_{(2)^2}\xrightarrow{f(x)} \F_{(2^2)^2}\xrightarrow{g(x)} \F_{((2^2)^2)^2}\xrightarrow{h(x)} \F_{(((2^2)^2)^2)^2} $.}}	\vspace{-0.1cm}
		\centering 
		\label{T:T1}
		
		\begin{tabular}{|c|c|c|c|c|}
			\hline
			\tiny  \bf  Finite Field & \tiny   \shortstack{\bf Normal Basis} & \tiny   \bf \shortstack{Normal element \\ as power of $ \omega $} & \tiny   \tiny \bf  \shortstack{Defining polynomial} \\
			\hline
			\tiny  $\F_{2^{16}}\cong \F_{(((2^2)^2)^2)^2} $ &\tiny \{$\delta, \delta^{256}$\}  &\tiny $ \delta= \omega^{45049} $  &\tiny $ h(x) = x^2 + x + \mu,  \mu = \beta + \lambda\gamma $  \\
			\hline			
			\tiny  $\F_{2^{8}}\cong \F_{((2^2)^2)^2}$ &\tiny \{$ \gamma, \gamma^{16} $\}  &\tiny $ \gamma = \omega^{14392} $  &\tiny $ g(x) = x^2 + x + \lambda , \lambda = \alpha^2\beta $  \\
\hline	
			\tiny  $\F_{2^{4}}\cong \F_{(2^2)^2}$ &\tiny \{$\beta, \beta^4$\}  &\tiny $ \beta = \omega^{4369}  $ &\tiny $ f(x) = x^2 + x + \alpha $ \\
\hline	
			\tiny  $\F_{2^{2}}\cong \F_{(2)^2}$ &\tiny \{$\alpha, \alpha^2$\}  &\tiny $ \alpha = \omega^{21845} $  &\tiny $ e(x) = x^2 + x + 1 $  \\
\hline				
		\end{tabular}\\
		~\\$ \omega $ is a root of polynomial $ p(x) $, used to construct the isomorphic field $\F_{2^{16}}$, $ \alpha $ a root of $ e(x) $, $ \beta $ a root of $ f(x) $, $ \gamma $ a root of $ g(x) $ and $ \delta $ a root of $ h(x) $.
	}	
	\vspace{-3mm}
\end{table}

\section{Composite Field Arithmetic and the Proposed Structures}
\label{CF}
In this section, we describe the arithmetic operations such as multiplication, square, and inversion in the composite fields. Also, the proposed circuits for the implementation of these operations are presented.

\subsection{Arithmetic operations in $\F_{2^2}$ and the proposed circuits}

Let $ A = a_0\alpha+a_1\alpha^2 $ and $ B = b_0\alpha+b_1\alpha^2 $, where $a_0, a_1, b_0, b_1\in \F_2$. A multiplication $ C = A\times B $ is computed as follows:\vspace{0.3cm} 

$M_2=A\times B = (a_0\alpha+a_1\alpha^2)(b_0\alpha + b_1\alpha^2)= a_0b_0\alpha2+(a_0b_1 + a_1b_0)(\alpha + \alpha^2) + a_1b_1\alpha= ((a_0\oplus a_1)(b_0\oplus b_1) \oplus a_0b_0)\alpha +((a_0 \oplus a_1)(b_0\oplus b_1) \oplus a_1b_1)\alpha^2= c_0\alpha + c_1\alpha^2$\vspace{0.3cm} 

The final equation of $c_1$ and $c_0$ after more simplification are as follows:\vspace{0.3cm} 

$c_0=(a_0\oplus a_1)(b_0\oplus b_1)\oplus a_0b_0=a_0b_0\oplus a_0b_1\oplus a_1b_1\oplus a_0b_0\oplus a_0b_0=a_0b_0\oplus a_0b_1\oplus a_1b_1=a_0b_1\oplus a_1(b_1\oplus b_0),$\vspace{0.1cm} 

$c_1=(a_0\oplus a_1)(b_0\oplus b_1)\oplus a_1b_1=a_0b_0\oplus a_0b_1\oplus a_1b_0\oplus a_1b_1\oplus a_1b_1=a_0b_1\oplus a_1b_0\oplus a_0b_0=a_1b_0\oplus a_0(b_1\oplus b_0).$\vspace{0.3cm} 

\begin{figure}[!t]
	\centering
	\includegraphics[scale=0.45]{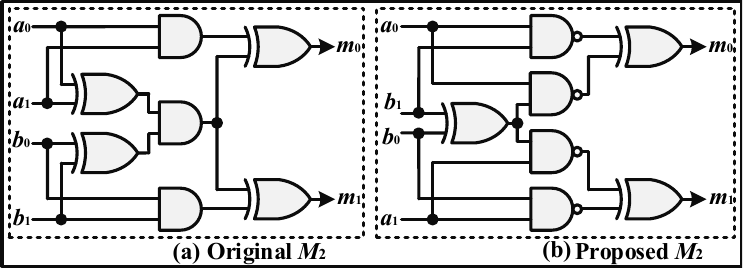}
	\caption{The original (a) and the proposed structure (b) of multiplication on the field $\F_{2^2}$.}
	\label{fig:FM2}
	\vspace{-6mm}
\end{figure}

Fig.\ref{fig:FM2} (a) and (b) show the original and the proposed structure of multiplication on the field $\F_{2^2}$ ($M_2$), respectively. In the proposed structure, the number of XOR gates is reduced.

For a non-zero element $A \in \F_{2^2}$, the square of $A$ is computed as follows:\vspace{0.3cm} 

$S_2=A^2= (a_0\alpha + a_1\alpha^2)^2= a_0\alpha^2+ a_1\alpha^4= a_1\alpha + a_0\alpha^2= s_0\alpha + s_1\alpha^2.$\vspace{0.3cm} 

Here, the inversion of $A \in \F_{2^2}$, i.e. $A^{-1}$ is equal to $A^2$. Moreover, the multiplications of $A \in \F_{2^2}$ by $\alpha$ and  $\alpha^2$ are performed as follows:\vspace{0.3cm} 

$ \alpha A = a_0\alpha^2+ a_1(\alpha + \alpha^2) = a_1\alpha + (a_0 \oplus a_1)\alpha^2 $\vspace{0.1cm}

$ \alpha^2A = a_0(\alpha + \alpha^2) + a_1\alpha = (a_0 \oplus a_1)\alpha + a_0\alpha^2.$\vspace{0.1cm}

\subsection{Arithmetic operations in $\F_{(2^2)^2}$ and the proposed circuits}

Let $ A=a_l\beta+a_h\beta^4 $ and $ B=b_l\beta+b_h\beta^4 $, where $ a_l, a_h, b_l, b_h \in \F_{2^2}$. A multiplication $ C = A\times B $ in $\F_{(2^2)^2}$ is computed as follows:\vspace{0.3cm}

$ M_4=A\times B = (a_l\beta + a_h\beta^4)(b_l\beta + b_h\beta^4)= a_lb_l\beta^2+ (a_lb_h + a_hb_l)\beta^5+ a_hb_1\beta^8= [(a_l \oplus a_h)(b_l \oplus b_h)\alpha \oplus a_lb_l]\beta +[(a_l \oplus a_h)(b_l \oplus b_h)\alpha \oplus a_hb_h]\beta^4= c_l\beta + c_h\beta^4= C. $\vspace{0.3cm}

\begin{figure}[!t]
	\centering
	\includegraphics[scale=0.3]{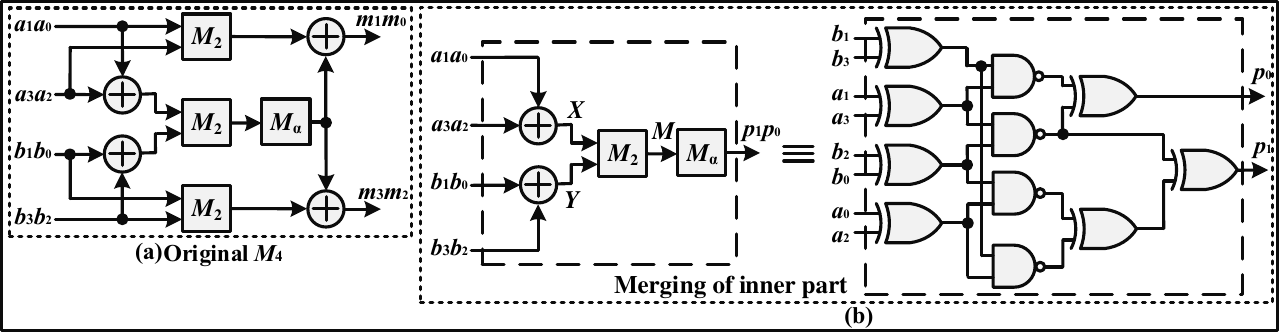}
	\caption{The multiplication in $\F_{(2^2)^2}$ with merging inner sub-blocks. The AND gates are converted to the NAND gate based on ($f\oplus e$)=($f'\oplus e'$).}
	\label{fig:FM4}
	\vspace{-6mm}
\end{figure}

The multiplication in $\F_{(2^2)^2}$ is shown in Fig.\ref{fig:FM4}. In the multiplication in $\F_{(2^2)^2}$ we merge the four blocks in the dot frame in Fig. \ref{fig:FM4} (b). For this propose, let we have $A=[a_3, a_2, a_1, a_0]$, and $B=[b_3, b_2, b_1, b_0]$, for two addition operations we have $X=[x_1,x_0]=[a_{1}\oplus a_{3}, a_{0}\oplus a_{2}]$ and $Y=[y_1,y_0]=[b_{1}\oplus b_{3}, b_{0}\oplus b_{2}]$. The output of the multiplication $M_2$, $ M=[m_1,m_0] $ is computed as follows:\vspace{0.3cm}

$m_0=x_0y_1\oplus x_1(y_1\oplus y_0)$, $m_1=x_1y_0\oplus x_0(y_1\oplus y_0)$.\vspace{0.3cm}

Also, the output of the $\alpha M$, $P=[p_1,p_0]$ block is computed as follows:\vspace{0.3cm}

$p_0=[(a_{1}\oplus a_{3})(b_{0}\oplus b_{2})]\oplus[(a_{0}\oplus a_{2})(b_{1}\oplus b_{3}\oplus b_{2} \oplus b_{0})]=(a_{1}\oplus a_{3})(b_{0}\oplus b_{2})\oplus (a_{0}\oplus a_{2})(b_{1}\oplus b_{3})\oplus(a_{0}\oplus a_{2})(b_{2}\oplus b_{0}),$\vspace{0.3cm}

$p_1=(a_{1}\oplus a_{3})(b_{0}\oplus b_{2})\oplus (a_{0}\oplus a_{2})(b_{1}\oplus b_{3}\oplus b_{2} \oplus b_{0})\oplus (a_{0}\oplus a_{2})(b_{1}\oplus b_{3})\oplus (a_{1}\oplus a_{3})(b_{1}\oplus b_{3}\oplus b_{2} \oplus b_{0})=(b_{1}\oplus b_{3}\oplus b_{2} \oplus b_{0})((a_{0}\oplus a_{2})\oplus (a_{1}\oplus a_{3}))\oplus(a_{1}\oplus a_{3})(b_{0}\oplus b_{2})\oplus (a_{0}\oplus a_{2})(b_{1}\oplus b_{3})=[(b_{1}\oplus b_{3})(a_{0}\oplus a_{2})]\oplus (b_{1}\oplus b_{3})(a_{1}\oplus a_{3})\oplus (b_{2}\oplus b_{0})(a_{0}\oplus a_{2})\oplus [(b_{2}\oplus b_{0})(a_{1}\oplus a_{3})]\oplus [(b_{1}\oplus b_{3})(a_{0}\oplus a_{2})]\oplus [(b_{2}\oplus b_{0})(a_{1}\oplus a_{3})]$=$(b_{1}\oplus b_{3})(a_{1}\oplus a_{3})\oplus (b_{2}\oplus b_{0})(a_{0}\oplus a_{2}).$\vspace{0.3cm}

For a non-zero element $A \in \F_{(2^2)^2}$, the square of $ A $ is computed as follows:\vspace{0.3cm}

$ A^2=(a_l\beta + a_h\beta^4)2= a^2_l\beta^2+ a^2_h\beta^8= a^2_l[(\alpha + 1)\beta + \alpha\beta^4] + a^2_h[\alpha\beta + (\alpha + 1)\beta^4]=[(a^2_l \oplus a^2_h)\alpha \oplus a^2_l]\beta + [(a^2_l \oplus a^2_h)\alpha \oplus a^2_h]\beta^4= s_0\beta + s_1\beta^4=S. $\vspace{0.3cm}

\begin{figure}[!t]
	\centering
	\includegraphics[scale=0.4]{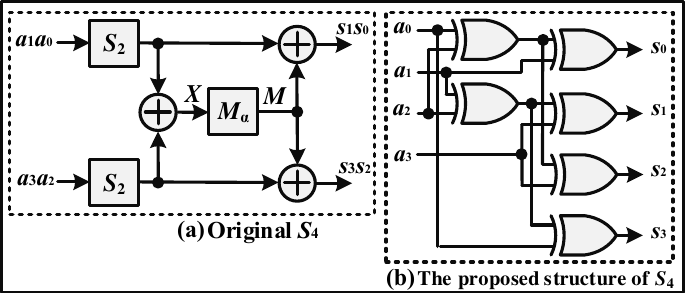}
	\caption{The the original circuit for the square of $ A $ over $ \F_{(2^2)^2} $ (a) the proposed implementation of the square (b).}
	\label{fig:FS4}
	\vspace{-6mm}
\end{figure}

Fig. \ref{fig:FS4} (a) shows the original circuit for the square of $ A $ over $ \F_{(2^2)^2} $. This circuit can be optimized to reduce gate count. Let $A=[a_3, a_2, a_1, a_0]$, the square $S_4=[s_3, s_2, s_1, s_0]$ is computed as follows:\vspace{0.3cm}

$ X=[a_{0}\oplus a_{2}, a_{1}\oplus a_{3}] $, $ M=[a_{0}\oplus a_{1}\oplus a_{2}\oplus a_{3}, a_{0}\oplus a_{2}] $\vspace{0.1cm}

$ s_{0}=a_{0}\oplus a_{2}\oplus a_{1},$\vspace{0.1cm}

$ s_{1}=a_{0}\oplus a_{2}\oplus a_{1}\oplus a_{3}\oplus a_{0}=a_{2}\oplus a_{1}\oplus a_{11},$\vspace{0.1cm}

$ s_{2}=a_{0}\oplus a_{2}\oplus a_{3},$\vspace{0.1cm}

$ s_{3}=a_{0}\oplus a_{2}\oplus a_{1}\oplus a_{3}\oplus a_{2}=a_{0}\oplus a_{1}\oplus a_{3}.$\vspace{0.3cm}

The proposed implementation of the square of $ A $ over $ \F_{(2^2)^2} $ is shown in Fig. \ref{fig:FS4} (b). As seen from the figure, the gate count of the proposed circuit is 39\% reduced than that of the original circuit. 

Letting $ A $ be a non-zero element in $ \F_{(2^2)^2} $, the inverse of $ A $, denoted by $ I_4 $, can be calculated by the Itoh-Tsujii algorithm (ITA) as follows:\vspace{0.3cm}

$I_4=A^{-1}= (AA^4)^{-1}A^4=
((a_l\beta + a_h\beta^4)(a_l\beta + a_h\beta^4)^4)^{-1}(a_l\beta + a_h\beta^4)^4=
 \cdots=
 ((a_l\beta + a_h\beta^4)(a_h\beta + a_l\beta^4))^{-1}(a_h\beta + a_l\beta^4)=((a_l \oplus a_h)^2\alpha \oplus a_la_h)^{-1}(a_1\beta + a_l\beta^4)=i_l\beta + i_h\beta^4. $\vspace{0.3cm}

\begin{figure}[!t]
	\centering
	\includegraphics[scale=0.35]{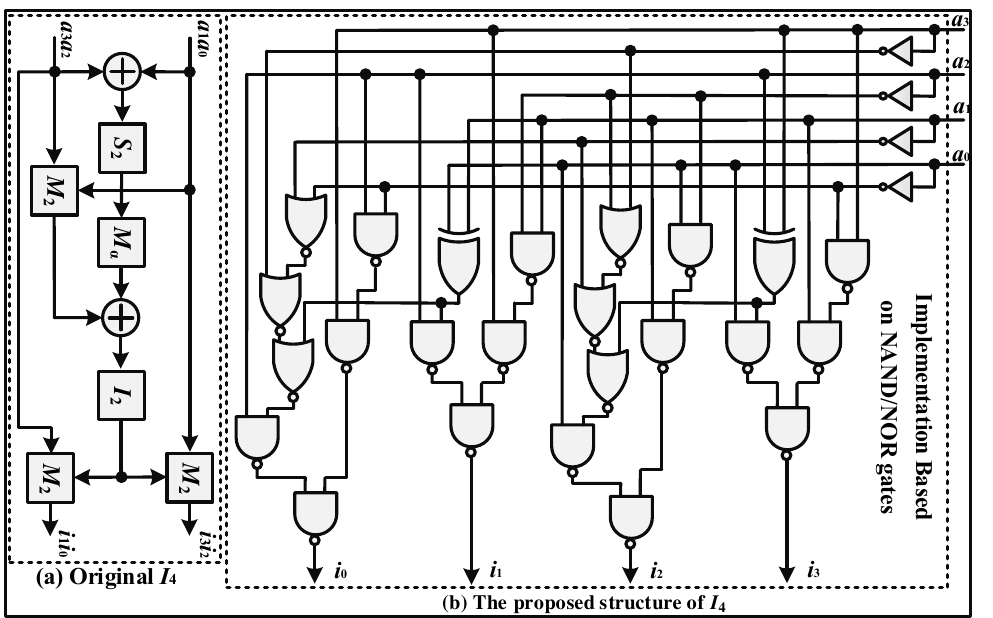}
	\caption{The original circuit for the inverse of $ A $ over $ \F_{(2^2)^2} $ (a) the proposed implementation of the inverse (b).}
	\label{fig:FI4}
	\vspace{-6mm}
\end{figure}

Fig. \ref{fig:FI4} (a) shows the original circuit for the inverse of $ A $ over $ \F_{(2^2)^2} $. To optimize the inverter circuit, we consider its input and output as a 4-variable truth table. By simplifying its outputs based on the Karnaugh map, a more optimal circuit for the inverter is obtained. Let $A=[a_3, a_2, a_1, a_0]$, the inverse $I_4=[i_3, i_2, i_1, i_0]$ based on Karnaugh map is computed as follows:\vspace{0.3cm}

$ i_{0}=a_{3}(a_{2}'+a_{0})+a_{2}(a_{1}\odot a_{0})(a_{3}'+a_{0}a_{1})$\vspace{0.1cm}

$ i_{1}=a_{3}(a_{1}'+a_{2})+a_{2}(a_{1}\oplus a_{0}),$\vspace{0.1cm}

$ i_{2}=a_{1}(a_{2}+a_{0}')+a_{0}(a_{3}\odot a_{2})(a_{2}a_{3}+a_{1}')$\vspace{0.1cm}

$ i_{3}=a_{1}(a_{3}'+a_{0})+a_{0}(a_{3}\oplus a_{2}).$\vspace{0.3cm}

The proposed implementation of the inverse of $ A $ over $ \F_{(2^2)^2} $ is shown in Fig. \ref{fig:FI4} (b). As seen from the figure, the gate count of the proposed circuit is 12\% reduced than that of the original circuit. 

The hardware results of the proposed structure of inversion and other works over $ \F_{(2^2)^2}$ is shown in Table \ref{T:T4}. The area and CPD for the proposed structure are equal to (2 2-input XOR/XNOR gates, 22 2-input NAND/NOR gates, and 4 NOT gates) and $3T_{NO}+2T_{NA}+T_N$, respectively. To reduce area and delay, in the proposed structure, the main part of the logic gates is implemented based on the NAND and NOR gates. The proposed structure of inversion over $\F_{2^4}$ has acceptable the area and delay parameters compared to other works.


\begin{table}[!t]
	\centering
	\captionsetup{justification=centering}
	{\scriptsize	
		\caption{Hardware complexity of the proposed structure of inversion in $ \F_{(2^2)^2}$ and other works.}\vspace{-0.3cm}
		\centering 
		\label{T:T4}
		
		\begin{tabular}{|c|c|c|c|c|c|}
			\hline
			\tiny  \bf Works & \tiny \bf \shortstack{XOR/\\XNOR} & \tiny \shortstack{AND/\\OR}  & \tiny \bf \shortstack{NAND/\\NOR} & \tiny \bf NOT&\tiny \bf CPD\\
			\hline		
			\tiny  \cite{7_4} &\tiny  14 &\tiny  8 &\tiny --- &\tiny --- &\tiny $3T_X+2T_A$\\
			\hline
			\tiny  \cite{12H} &\tiny  4 &\tiny  --- &\tiny 14 &\tiny 4 &\tiny $T_X+2T_{NO}+T_N$\\
			\hline
			\tiny  \cite{30} &\tiny  4 &\tiny  --- &\tiny 12 (+4 NAND3) &\tiny 4 &\tiny $3T_{NA}+T_N$\\
			\hline
			\tiny  \cite{41} &\tiny  10 &\tiny  12 &\tiny --- &\tiny 4 &\tiny $2T_X+2T_{A}+T_N$\\
			\hline
			\tiny  \cite{BRI} &\tiny  8  &\tiny --- &\tiny 8 &\tiny 3 &\tiny $2T_X+2T_{NA}$  \\
			\hline 
			\tiny  \cite{BRI2} &\tiny  4  &\tiny --- &\tiny 21 &\tiny 4 &\tiny $T_{X}+2T_{NA}+T_{NO}+T_N$  \\
			\hline			
			\tiny  This work &\tiny  2  &\tiny --- &\tiny 22 &\tiny 4 &\tiny $3T_{NO}+2T_{NA}+T_N$  \\
			\hline	
		\end{tabular}\\
		~\\
		$T_A$, $T_{NA}$, $T_X$, $T_{NO}$, $T_{N}$ denote the time delay of a 2-input AND gate, 2-input NAND gate, 2-input XOR gate, 2-input NOR gate, and NOT gate, respectively.
	}	
	\vspace{-6mm}
\end{table}

The multiplications of $A \in \F_{(2^2)^2}$ by $ \lambda $, $ \lambda^2 $, $ \beta$, and $ \alpha\beta $ and also the proposed computations for the hardware implementation are performed as follows:\vspace{0.3cm}

$ \lambda A = \alpha^2\beta A= a_l\alpha^2[(\alpha + 1)\beta + \alpha\beta^4] + a_h\alpha^2(\alpha\beta + \alpha\beta^4)= (a_l\alpha \oplus a_h)\beta + (a_l \oplus a_h)\beta^4, $\vspace{0.1cm}

$ M_{0}=a_{1}$, $ M_{1}=a_{1}\oplus a_{0}$\vspace{0.1cm}

$ K_{0}=M_{0}\oplus a_{2}=a_{1}\oplus a_{2}$, $ K_{1}=M_{1}\oplus a_{3}=a_{1}\oplus a_{0}\oplus a_{3}$, $ K_{2}=a_{0}\oplus a_{2}$, $ K_{3}=a_{1}\oplus a_{3}$.\vspace{0.3cm}

$ \lambda^2A = \alpha\beta2A= a_0\alpha\beta^3+ a_1\alpha\beta^6= a_0\alpha(\beta + \alpha\beta^4) + a_1\alpha^2\beta= (a_0\alpha \oplus a_1\alpha^2)\beta + (a_0\alpha^2)\beta^4, $\vspace{0.1cm}

$ T_{0}=a_{1}$, $ T_{1}=a_{1}\oplus a_{0}$.\vspace{0.1cm}

$ L_{0}=a_{2}\oplus a_{3}$, $ L_{1}=a_{2}$\vspace{0.1cm}

$ K_{0}=a_{1}\oplus a_{2}\oplus a_{3}$, $ K_{1}=a_{1}\oplus a_{0}\oplus a_{2}$, $ K_{2}=T_1=a_{1}\oplus a_{0}$, $ K_{3}=T_0\oplus T_1=a_{1}\oplus a_{1}\oplus a_{0}=a_{0}$. \vspace{0.3cm}

$ \beta A = a_0[(\alpha + 1)\beta + \alpha\beta^4] + a_1(\alpha\beta + \alpha\beta^4)= [a_0 \oplus (a_0 \oplus a_1)\alpha]\beta + [(a_0 \oplus a_1)\alpha]\beta^4, $\vspace{0.1cm}

$ T_{0}=a_{0}\oplus a_{2}$, $ T_{1}=a_{1}\oplus a_{3}$.\vspace{0.1cm}

$ L_{0}=T_1=a_{1}\oplus a_{3}$, $ L_{1}=T_0\oplus T_1=a_{0}\oplus a_{2}\oplus a_{1}\oplus a_{3}$.\vspace{0.1cm}

$ K_{0}=a_{0}\oplus a_{1}\oplus a_{3}$, $ K_{1}=a_{1}\oplus a_{0}\oplus a_{2}\oplus a_{1}\oplus a_{3}=a_{1}\oplus a_{0}\oplus a_{2}\oplus a_{3}$, $ K_{2}=T_1=a_{1}\oplus a_{3}$, $ K_{3}=a_{0}\oplus a_{2}\oplus a_{1}\oplus a_{3}$.\vspace{0.3cm}

$ \alpha\beta A = a_0(\beta + \alpha^2\beta^4) + a_1(\alpha^2\beta + \alpha^2\beta^4)= (a_0 \oplus a_1\alpha^2)\beta + (a_0\alpha^2\oplus a_1\alpha^2)\beta^4. $\vspace{0.1cm}

$ T_{0}=a_{0}\oplus a_{1}$, $ T_{1}=a_{0}$.\vspace{0.1cm}

$ L_{0}=a_{2}\oplus a_{3}$, $ L_{1}=a_{2}$.\vspace{0.1cm}

$ K_{0}=a_{0}\oplus a_{1}\oplus a_{3}$, $ K_{1}=a_{1}\oplus a_{2}$, $ K_{2}=a_{0}\oplus a_{1}\oplus a_{2}\oplus a_{3}$, $ K_{3}=a_{2}\oplus a_{0}$. \vspace{0.3cm}

\begin{figure}[!t]
	\centering
	\includegraphics[scale=0.36]{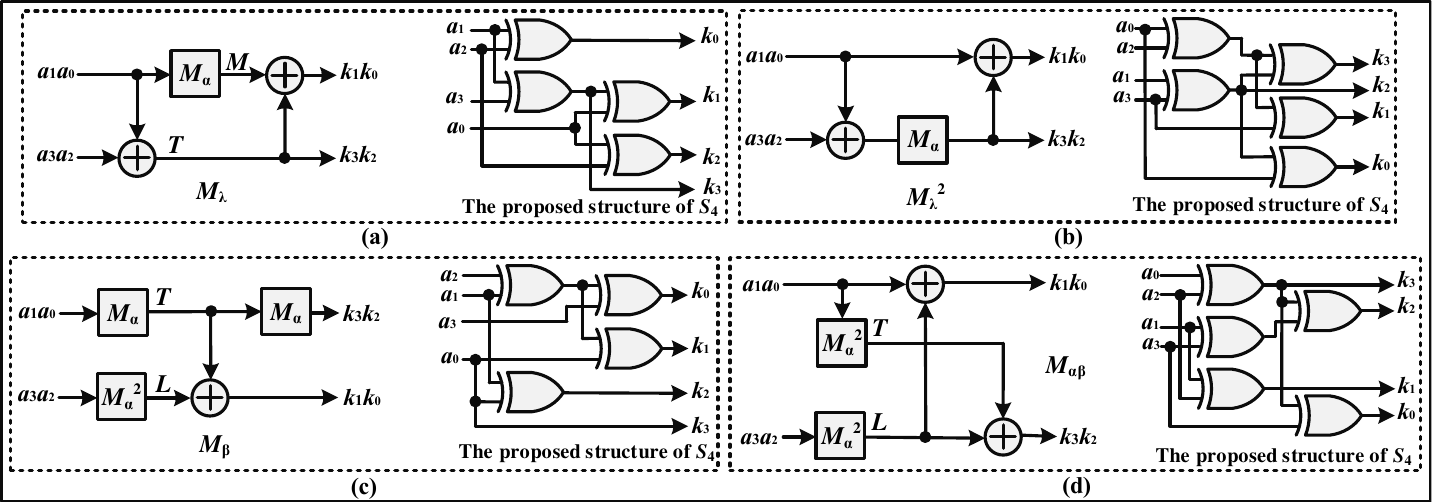}
	\caption{The proposed implementations of the multiplications of $A \in \F_{(2^2)^2}$ by $ \lambda $ (a), $ \lambda^2 $ (b), $ \beta $ (c) and $ \alpha\beta $ (d).}
	\label{fig:FMHA}
	\vspace{-6mm}
\end{figure}

The proposed implementations for the multiplications of $A \in \F_{(2^2)^2}$ by $ \lambda $, $ \lambda^2 $, $ \beta $ and $ \alpha\beta $ are shown in Figs. \ref{fig:FMHA} (a)-(d), respectively. As seen from the figures, the gate count of the proposed circuits are reduced than that of the original circuits. 

\subsection{Arithmetic operations in $\F_{((2^2)^2)^2}$ and the proposed circuits}

\begin{figure}[!t]
	\centering
	\includegraphics[scale=0.4]{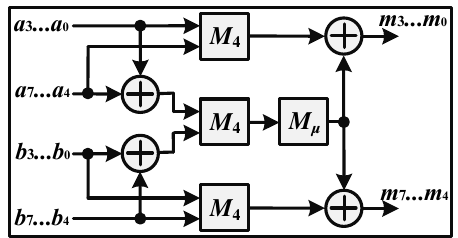}
	\caption{The structure of multiplication in $\F_{((2^2)^2)^2}$.}
	\label{fig:FM8}
	\vspace{-6mm}
\end{figure}

The structure of multiplication in $\F_{((2^2)^2)^2}$ is shown in Fig. \ref{fig:FM8}. Let $A = a_l\gamma+a_h\gamma^{16}$ and $B = b_l\gamma+b_h\gamma^{16}$, where $a_l, a_h, b_l, b_h  \in \F_{(2^2)^2}$. A multiplication $C = A\times B$ in $\F_{((2^2)^2)^2}$ is carried out as follows:\vspace{0.3cm}

$M_8=A\times B = (a_l\gamma + a_h\gamma^{16})(b_l\gamma + b_h\gamma^{16})= a_lb_l\gamma^2+ (a_lb_h + a_hb_l)\gamma{17}+ a_hb_h\gamma^{32}= [(a_l + a_h)(b_l + b_h)\gamma + a_lb_l]\gamma +[(a_l + a_h)(b_l + b_h)\gamma + a_hb_h]\gamma^{16}= c_l\gamma + c_h\gamma^{16}= C.$\vspace{0.3cm}

For a non-zero element $ A \in \F_{((2^2)^2)^2} $, the square of $A$ is calculated as follows:\vspace{0.3cm}

$A^2= (a_l\gamma + a_h\gamma^{16})2= a^2_l\gamma^2+ a^2_h\gamma^{32}= a^2_l[(\lambda + 1)\gamma + \lambda\gamma^{16}] + a^2_h[\lambda\gamma + (\lambda + 1)\gamma^{16}]= [(a^2_l + a^2_h)\lambda + a^2_l]\gamma + [(a^2_l + a^2_h)\lambda + a^2_h]\gamma^{16}= s_l\gamma + s_h\gamma^{16}= S_8.$\vspace{0.3cm}

\begin{figure}[!t]
	\centering
	\includegraphics[scale=0.4]{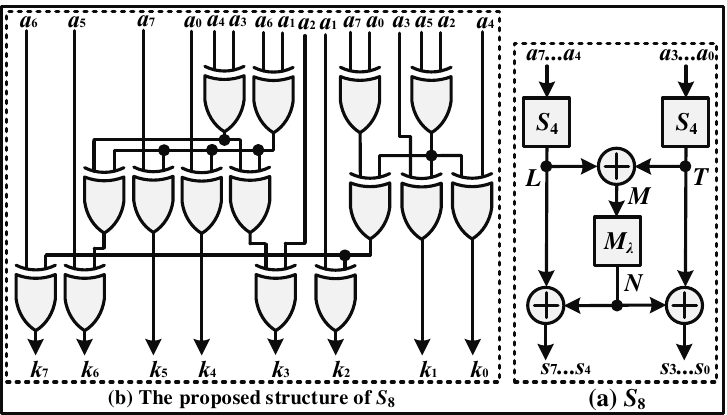}\vspace{-2mm}
	\caption{The original circuit for the square of $ A $ over $\F_{((2^2)^2)^2}$ (a) the proposed structure of the squaring (b).}
	\label{fig:FS8}
	\vspace{-6mm}
\end{figure}

Fig. \ref{fig:FS8} (a) shows the original circuit for the square of $ A $ over $\F_{((2^2)^2)^2}$. In the following, we present the optimized computations for implementation of this computation unit. Let $A=[a_7, a_6, a_5, a_4, a_3, a_2, a_1, a_0]$, the square result $S_8=[s_7, s_6, s_5, s_4, s_3, s_2, s_1, s_0]$ is computed as follows:\vspace{0.3cm}

$ T_{0}=a_{0}\oplus a_{2}\oplus a_{1}, T_{1}=a_{2}\oplus a_{1}\oplus a_{3}, T_{2}=a_{0}\oplus a_{2}\oplus a_{3}, T_{3}=a_{0}\oplus a_{1}\oplus a_{3}.$\vspace{0.1cm}

$ L_{0}=a_{4}\oplus a_{6}\oplus a_{5}, L_{1}=a_{6}\oplus a_{5}\oplus a_{7}, L_{2}=a_{4}\oplus a_{6}\oplus a_{7}, L_{3}=a_{4}\oplus a_{5}\oplus a_{7}.$\vspace{0.1cm}

$ M_{0}=T_{0}\oplus L_{0}, M_{1}=T_{1}\oplus L_{1}, M_{2}=T_{2}\oplus L_{2}, M_{3}=T_{3}\oplus L_{3}.$\vspace{0.1cm}

$ N_{0}=M_{1}\oplus M_{2}=(T_{1}\oplus L_{1}) \oplus (T_{2}\oplus L_{2}), N_{1}=M_{1}\oplus M_{0} \oplus M_{3}=(T_{1}\oplus L_{1}) \oplus (T_{0}\oplus L_{0})\oplus (T_{3}\oplus L_{3}), N_{2}=M_{0}\oplus M_{2}=(T_{0}\oplus L_{0}) \oplus (T_{2}\oplus L_{2}), N_{3}=M_{1}\oplus M_{3}=(T_{1}\oplus L_{1}) \oplus (T_{3}\oplus L_{3}).$\vspace{0.3cm}

Table \ref{T:T2} shows The optimized computations for implementation of the square of $ A $ over $\F_{((2^2)^2)^2}$. The proposed structure of the squaring in $\F_{((2^2)^2)^2}$ is shown in Fig. \ref{fig:FS8} (b). This circuit is compacted compare to the original circuit. The number of XOR gates of the proposed circuit is 65\% reduced than that of the original circuit. The critical path delay of the proposed and original structures of the squaring in $\F_{((2^2)^2)^2}$ is equal to 3$T_X$ and 7$T_X$, respectively, where $T_X$ is denote the time delay of a 2-input XOR gate.

\begin{table}[!t]
	\centering
	\captionsetup{justification=centering}
	{\scriptsize	
		\caption{{\scriptsize The optimized computations for the square of $ A $ over $\F_{((2^2)^2)^2}$.}}\vspace{-0.2cm}
		\centering 
		\label{T:T2}
		
		\begin{tabular}{|c|p{7cm}|}
			\hline
			\tiny  \bf $s_i$ & \tiny \bf Computations \\
			\hline		
			\tiny  $ s_{0}$ &\tiny $T_{1}\oplus N_{1}=(a_{2}\oplus a_{1}\oplus a_{3})\oplus (a_{2}\oplus a_{1}\oplus a_{3})\oplus(a_{6}\oplus a_{5}\oplus a_{7})\oplus (a_{0}\oplus a_{2}\oplus a_{1})\oplus(a_{4}\oplus a_{6}\oplus a_{5})\oplus(a_{0}\oplus a_{1}\oplus a_{3})\oplus(a_{4}\oplus a_{5}\oplus a_{7})=a_{2}\oplus a_{3}\oplus a_{5}$\\
			\hline	
			\tiny  $ s_{1}$ &\tiny $T_{0}\oplus N_{0}=a_{0}\oplus a_{2}\oplus a_{1}\oplus (a_{2}\oplus a_{1}\oplus a_{3})\oplus(a_{6}\oplus a_{5}\oplus a_{7})\oplus (a_{0}\oplus a_{2}\oplus a_{3})\oplus(a_{4}\oplus a_{6}\oplus a_{7})=a_{4}\oplus a_{2}\oplus a_{5}$\\
			\hline	
			\tiny  $ s_{2}$ &\tiny $T_{2}\oplus N_{2}=a_{0}\oplus a_{2}\oplus a_{3}\oplus (a_{0}\oplus a_{2}\oplus a_{1})\oplus(a_{4}\oplus a_{6}\oplus a_{5})\oplus (a_{0}\oplus a_{2}\oplus a_{3})\oplus(a_{4}\oplus a_{6}\oplus a_{7})=a_{1}\oplus a_{5}\oplus a_{0}\oplus a_{2}\oplus a_{7}$\\
			\hline	
			\tiny  $ s_{3}$ &\tiny $T_{3}\oplus N_{3}=a_{0}\oplus a_{1}\oplus a_{3}\oplus (a_{2}\oplus a_{1}\oplus a_{3})\oplus(a_{6}\oplus a_{5}\oplus a_{7})\oplus (a_{0}\oplus a_{1}\oplus a_{3})\oplus(a_{4}\oplus a_{5}\oplus a_{7})=a_{2}\oplus a_{1}\oplus a_{3}\oplus a_{6}\oplus a_{4}$\\
			\hline	
			\tiny  $ s_{4}$ &\tiny $L_{0}\oplus N_{0}=a_{4}\oplus a_{6}\oplus a_{5}\oplus (a_{2}\oplus a_{1}\oplus a_{3})\oplus(a_{6}\oplus a_{5}\oplus a_{7})\oplus (a_{0}\oplus a_{2}\oplus a_{3})\oplus(a_{4}\oplus a_{6}\oplus a_{7})=a_{1}\oplus a_{6}\oplus a_{0}$\\
			\hline	
			\tiny  $ s_{5}$ &\tiny $L_{1}\oplus N_{1}=(a_{6}\oplus a_{5}\oplus a_{7})\oplus (a_{2}\oplus a_{1}\oplus a_{3})\oplus(a_{6}\oplus a_{5}\oplus a_{7})\oplus (a_{0}\oplus a_{2}\oplus a_{1})\oplus(a_{4}\oplus a_{6}\oplus a_{5})\oplus(a_{0}\oplus a_{1}\oplus a_{3})\oplus(a_{4}\oplus a_{5}\oplus a_{7})=a_{6}\oplus a_{1}\oplus a_{7}$\\
			\hline	
			\tiny  $ s_{6}$ &\tiny $L_{2}\oplus N_{2}=a_{4}\oplus a_{6}\oplus a_{7}\oplus (a_{0}\oplus a_{2}\oplus a_{1})\oplus(a_{4}\oplus a_{6}\oplus a_{5})\oplus (a_{0}\oplus a_{2}\oplus a_{3})\oplus(a_{4}\oplus a_{6}\oplus a_{7})=a_{1}\oplus a_{5}\oplus a_{3}\oplus a_{4}\oplus a_{6}$\\
			\hline	
			\tiny  $ s_{7}$ &\tiny $L_{3}\oplus N_{3}=a_{4}\oplus a_{5}\oplus a_{7}\oplus (a_{2}\oplus a_{1}\oplus a_{3})\oplus(a_{6}\oplus a_{5}\oplus a_{7})\oplus (a_{0}\oplus a_{1}\oplus a_{3})\oplus(a_{4}\oplus a_{5}\oplus a_{7})=a_{2}\oplus a_{6}\oplus a_{5}\oplus a_{7}\oplus a_{0}$\\
		\hline		
		\end{tabular}\\
		~\\
	}	
	\vspace{-2.5mm}
\end{table}

Letting $ A $ be a non-zero element in $\F_{((2^2)^2)^2}$, the inverse of $A$, denoted by $I_8$, can be computed by the ITA method as follows:\vspace{0.3cm}

$I_8=A^{-1}=(AA^{16})^{-1}A^{16}= ((a_l\gamma + a_h\gamma^{16})(a_l\gamma + a_h\gamma^{16})^{16})^{-1}(a_l\gamma + a_h\gamma^{16})^{16}=[(a_l\gamma + a_h\gamma^{16})(a_h\gamma + a_l\gamma^{16})]^{-1}(a_h\gamma + a_l\gamma^{16})=((a_l\gamma + a_h\gamma^{16})(a_l\gamma^{16}+ a_h\gamma^{256}))^{-1}(a_l\gamma^{16}+ a_h\gamma^{256})=((a_l\gamma + a_h\gamma^{16})(a_h\gamma + a_l\gamma^{16}))^{-1}(a_h\gamma + a_l\gamma^{16})=((a_l + a_h)2\lambda + a_la_h)^{-1}(a_h\gamma + a_l\gamma^{16})= i_l\gamma + i_h\gamma^{16}.$\vspace{0.3cm}

\begin{figure}[!t]
	\centering
	\includegraphics[scale=0.4]{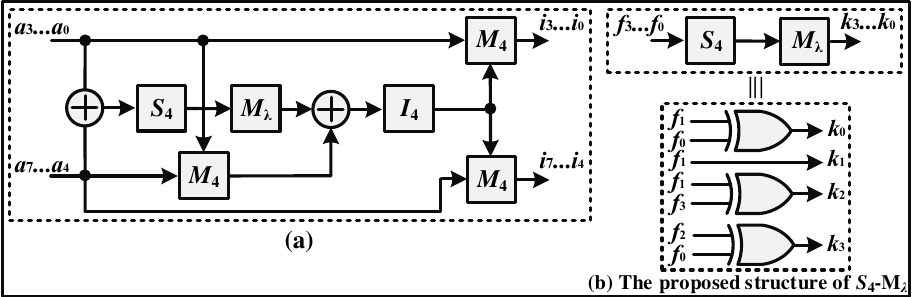}
	\caption{The original circuit for the inverse of $ A $ over $\F_{((2^2)^2)^2}$ (a) the merging inner sub-blocks $S_4$ and $M_\lambda$ (b).}
	\label{fig:FI8}
	\vspace{-6mm}
\end{figure}

Fig. \ref{fig:FI8} (a) shows the original circuit for the inverse of $ A $ over $\F_{((2^2)^2)^2}$. In the following, we merge the two blocks $S_4$ and $M_\lambda$ to reduce the hardware consumption (Fig. \ref{fig:FI8} (b)). Merging the two blocks $S_4$ and $M_\lambda$ is computed as follows:\vspace{0.3cm}

$ T_{0}=f_{0}\oplus f_{2}\oplus f_{1}, T_{1}=f_{2}\oplus f_{1}\oplus f_{3}, T_{2}=f_{0}\oplus f_{2}\oplus f_{3}, T_{3}=f_{0}\oplus f_{1}\oplus f_{3}.$\vspace{0.1cm}

$ k_{0}=T_{1}\oplus T_{2}=f_{2}\oplus f_{1}\oplus f_{3}\oplus f_{0}\oplus f_{2}\oplus f_{3}=f_{1}\oplus f_{0},$\vspace{0.1cm}

$ k_{1}=T_{1}\oplus T_{0}\oplus T_{3}=f_{2}\oplus f_{1}\oplus f_{3}\oplus f_{0}\oplus f_{2}\oplus f_{1}\oplus f_{0}\oplus f_{1}\oplus f_{3}=f_{1},$\vspace{0.1cm}

$ k_{2}=T_{0}\oplus T_{2}=f_{0}\oplus f_{2}\oplus f_{1}\oplus f_{0}\oplus f_{2}\oplus f_{3}=f_{1}\oplus f_{3},$\vspace{0.1cm}

$ k_{3}=T_{1}\oplus T_{3}=f_{2}\oplus f_{1}\oplus f_{3}\oplus f_{0}\oplus f_{1}\oplus f_{3}=f_{2}\oplus f_{0}.$\vspace{0.3cm}

Multiplication of $ A \in \F_{((2^2)^2)^2} $ by $ \mu $ is one of the important operation. This operation is computed as follows: \vspace{0.3cm}

$\mu A = (\beta + \lambda\gamma)(a_l\gamma+ a_h\gamma^{16})= a_l\beta\gamma + a_h\beta\gamma^{16}+a_l\lambda\gamma^2+ a_h\lambda\gamma^{17}= [a_l(\alpha\beta) + (a_l + a_h)\lambda^2]\gamma + [a_h\beta + (a_l + a_h)\lambda^2]\gamma^{16}$\vspace{0.3cm}

\begin{figure}[!t]
	\centering
	\includegraphics[scale=0.4]{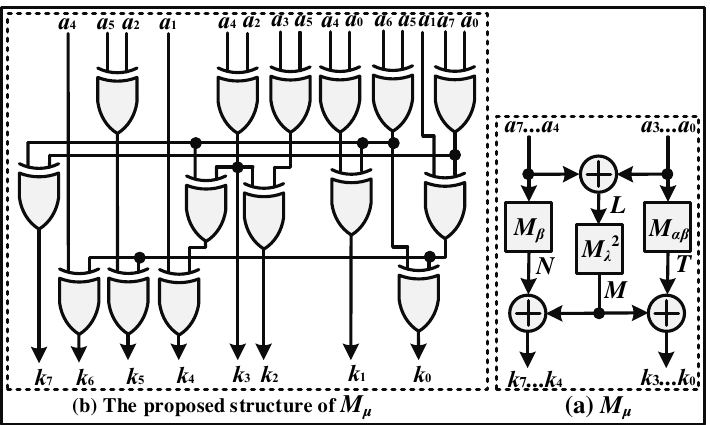}
	\caption{The original circuit for the multiplication of $ A \in \F_{((2^2)^2)^2} $ by $ \mu $ (a), the proposed structure of the multiplication by $ \mu $ (b).}
	\label{fig:FMU}
	\vspace{-6mm}
\end{figure}

Fig. \ref{fig:FMU} (a) shows the original circuit for the multiplication of $ A \in \F_{((2^2)^2)^2} $ by $ \mu $. The optimized computations for the $M_{\mu}$ is presented as follows:\vspace{0.3cm}

$ L_{0}=a_{0}\oplus a_{4}, L_{1}=a_{1}\oplus a_{5}, L_{2}=a_{2}\oplus a_{6}, L_{3}=a_{3}\oplus a_{7}, T_{0}=a_{0}\oplus a_{2}\oplus a_{3}, T_{1}=a_{1}\oplus a_{2}, T_{2}=a_{0}\oplus a_{1}\oplus a_{2}\oplus a_{3}, T_{3}=a_{2}\oplus a_{0}.$\vspace{0.1cm}

$ N_{0}=a_{4}\oplus a_{5}\oplus a_{7}, N_{1}=a_{4}\oplus a_{6}\oplus a_{7}, N_{2}=a_{5}\oplus a_{7}, N_{3}=a_{4}\oplus a_{6}\oplus a_{5}\oplus a_{7}, M_{0}=L_{1}\oplus L_{2}\oplus L_{3}=a_{1}\oplus a_{5}\oplus a_{2}\oplus a_{6}\oplus a_{3}\oplus a_{7}, M_{1}=L_{1}\oplus L_{2}\oplus L_{0}=a_{1}\oplus a_{11}\oplus a_{2}\oplus a_{6}\oplus a_{0}\oplus a_{4}, M_{2}=L_{1}\oplus L_{0}=a_{1}\oplus a_{5}\oplus a_{0}\oplus a_{4}, M_{3}=L_{0}=a_{0}\oplus a_{4}.$\vspace{0.1cm}

$ k_{0}=T_{0}\oplus M_{0}=a_{0}\oplus a_{2}\oplus a_{3}\oplus a_{1}\oplus a_{5}\oplus a_{2}\oplus a_{6}\oplus a_{3}\oplus a_{7}=a_{0}\oplus a_{1}\oplus a_{5}\oplus a_{6}\oplus a_{7},$\vspace{0.1cm}

$ k_{1}=T_{1}\oplus M_{1}=a_{1}\oplus a_{2}\oplus a_{1}\oplus a_{5}\oplus a_{2}\oplus a_{6}\oplus(a_{0}\oplus a_{4}=a_{5}\oplus a_{6}\oplus a_{0}\oplus a_{4}$,\vspace{0.1cm}

$ k_{2}=T_{2}\oplus M_{2}=a_{0}\oplus a_{1}\oplus a_{2}\oplus a_{3}\oplus a_{1}\oplus a_{5}\oplus a_{0}\oplus a_{4}=a_{2}\oplus a_{3}\oplus a_{5}\oplus a_{4},$\vspace{0.1cm}

$ k_{3}=T_{3}\oplus M_{3}=a_{2}\oplus a_{0}\oplus a_{0}\oplus a_{4}=a_{2}\oplus a_{4},$\vspace{0.1cm}

$ k_{4}=N_{0}\oplus M_{0}=a_{4}\oplus a_{5}\oplus a_{7}\oplus a_{1}\oplus a_{5}\oplus a_{2}\oplus a_{6}\oplus a_{3}\oplus a_{7}=a_{4}\oplus a_{01}\oplus a_{5}\oplus a_{2}\oplus a_{0}$,\vspace{0.1cm}

$ k_{5}=N_{1}\oplus M_{1}=a_{4}\oplus a_{6}\oplus a_{7}\oplus a_{1}\oplus a_{5}\oplus a_{2}\oplus a_{6}\oplus a_{0}\oplus a_{4}=a_{7}\oplus a_{1}\oplus a_{5}\oplus a_{2}\oplus a_{0}$,\vspace{0.1cm}

$ k_{6}=N_{2}\oplus M_{2}=a_{5}\oplus a_{7}\oplus a_{1}\oplus a_{5}\oplus a_{0}\oplus a_{4}=a_{7}\oplus a_{1}\oplus a_{0}\oplus a_{4}$,\vspace{0.1cm}

$ k_{7}=N_{3}\oplus M_{3}=a_{4}\oplus a_{6}\oplus a_{5}\oplus a_{7}\oplus a_{0}\oplus a_{4}=a_{6}\oplus a_{5}\oplus a_{7}\oplus a_{0}$.\vspace{0.3cm}

The proposed structure of the multiplication by $ \mu $ in $\F_{((2^2)^2)^2}$ is shown in Fig. \ref{fig:FMU} (b). The hardware resources of this circuit is reduced than those of the original circuit. The proposed circuit features a 46\% reduction in the number of XOR gates compared to the original circuit. In terms of critical path delay, the proposed structure for multiplication by $ \mu $ in $\F_{((2^2)^2)^2}$ has a delay of $3T_X$, while the original structure has a delay of $5T_X$, where $T_X$ represents the time delay of a 2-input XOR gate.

\subsection{Inversion operation in $\F_{(((2^2)^2)^2)^2}$ and the proposed circuits}

%
%

%

The proposed S-box structure is based on the field inversion in the finite field $\F_{(((2^2)^2)^2)^2}$. Therefore, in this section, the structure of the inversion unit in this field is explained. Letting $ A $ be a non-zero element in $\F_{(((2^2)^2)^2)^2}$, the inverse of $ A $, denoted by $ I_{16} $, can be computed by the ITA method as follows:\vspace{0.1cm}

$ I_{16}=A^{-1}= (AA^{256})^{-1}A^{256}= ((a_l\delta + a_h\delta^{256})(a_l\delta + a_h\delta^{256})^{256})^{-1}(a_l\delta + a_h\delta^{256})^{256}=((a_l\delta + a_h\delta^{256})(a_l\delta^{256}+ a_h\delta^{65536}))^{-1}(a_l\delta^{256}+ a_h\delta^{65536})=((a_l\delta + a_h\delta^{256})(a_h\delta + a_l\delta^{256}))^{-1}(a_h\delta + a_l\delta^{256})= ((a_l + a_h)^2\mu + a_la_h)^{-1}(a_h\delta + a_l\delta^{256})= i_l\delta + i_h\delta^{256}. $\vspace{0.3cm}

\begin{figure}[!t]
	\centering
	\includegraphics[scale=0.4]{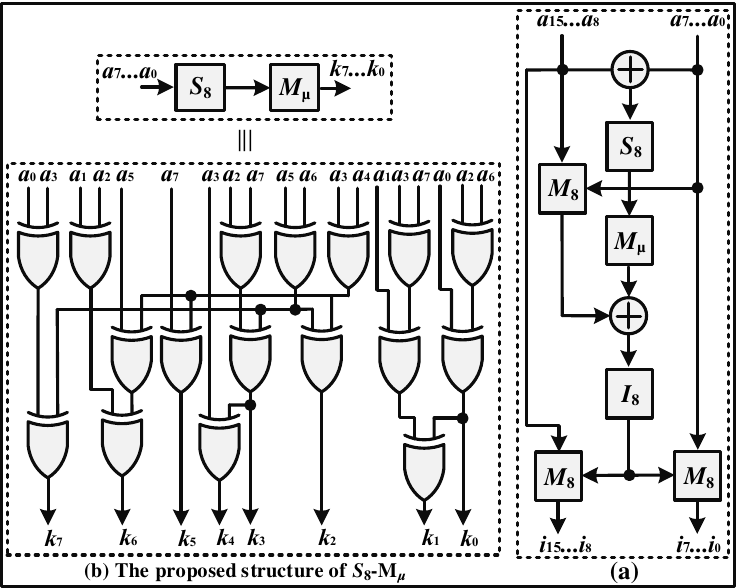}
	\caption{The original circuit for the inverse of $ A $ over $\F_{(((2^2)^2)^2)^2}$ (a) the merging inner sub-blocks $S_8$ and $M_\mu$ (b).}
	\label{fig:FI16}
	\vspace{-6mm}
\end{figure}

Fig. \ref{fig:FI16} (a) shows the original circuit for the inverse of $ A $ over $\F_{(((2^2)^2)^2)^2}$. In the inversion operation, we have two blocks $S_8$ and $M_\mu$ so that these blocks can be combined to reduce hardware resources. In the following, we merge the two blocks $S_8$ and $M_\mu$ to reduce the hardware consumption (Fig. \ref{fig:FI16} (b)). Merging the two blocks $S_8$ and $M_\mu$ is computed as follows:\vspace{0.1cm}

$ s_{0}=a_{4}\oplus a_{2}\oplus a_{5}$, $ s_{1}=a_{2}\oplus a_{3}\oplus a_{5}$, $ s_{2}=a_{1}\oplus a_{5}\oplus a_{0}\oplus a_{2}\oplus a_{7}$, $ s_{3}=a_{2}\oplus a_{1}\oplus a_{3}\oplus a_{6}\oplus a_{4}$, $ s_{4}=a_{1}\oplus a_{6}\oplus a_{0}$, $ s_{5}=a_{6}\oplus a_{1}\oplus a_{7}$, $ s_{6}=a_{1}\oplus a_{5}\oplus a_{3}\oplus a_{4}\oplus a_{6}$, $ s_{7}=a_{2}\oplus a_{6}\oplus a_{5}\oplus a_{7}\oplus a_{0}$. \vspace{0.3cm}

The final computations of the merging two blocks $S_8$ and $M_\mu$ is shown in Table \ref{T:T3}. Fig. \ref{fig:FI16} (b) shows the merging two blocks $S_8$ and $M_\mu$. The number of XOR gates for the proposed circuit merging two blocks $S_8$ and $M_\mu$ and original circuit are 17 and 71, respectively. The proposed circuit achieves a 76\% decrease in the number of XOR gates relative to the original circuit. Regarding critical path delay, the proposed design for the merging two blocks $S_8$ and $M_\mu$ exhibits a delay of $3T_X$, whereas the original design has a delay of $12T_X$.

\begin{table}[!t]
	\centering
	\captionsetup{justification=centering}
	{\scriptsize	
		\caption{The final computations of the merging two blocks $S_8$ and $M_\mu$.}\vspace{-0.3cm}
		\centering 
		\label{T:T3}
		
		\begin{tabular}{|c|p{7cm}|}
			\hline
			\tiny  \bf $k_i$ & \tiny \bf Computations \\
			\hline		
			\tiny  $ k_{0}$ &\tiny $s_{0}\oplus s_{1}\oplus s_{5}\oplus s_{6}\oplus s_{7}=a_{4}\oplus a_{2}\oplus a_{5}\oplus a_{2}\oplus a_{3}\oplus a_{5}\oplus a_{6}\oplus a_{1}\oplus a_{7}\oplus a_{1}\oplus a_{5}\oplus a_{3}\oplus a_{4}\oplus a_{6}\oplus a_{2}\oplus a_{6}\oplus a_{5}\oplus a_{7}\oplus a_{0}=a_{2}\oplus a_{6}\oplus a_{0}$\\
			\hline	
			\tiny  $ k_{1}$ &\tiny $s_{5}\oplus s_{6}\oplus s_{0}\oplus s_{4}=a_{6}\oplus a_{1}\oplus a_{7}\oplus a_{1}\oplus a_{5}\oplus a_{3}\oplus a_{4}\oplus a_{6}\oplus a_{4}\oplus a_{2}\oplus a_{5}\oplus a_{1}\oplus a_{6}\oplus a_{0}=a_{7}\oplus a_{3}\oplus a_{2}\oplus a_{1}\oplus a_{6}\oplus a_{0}$\\
			\hline	
			\tiny  $ k_{2}$ &\tiny $s_{2}\oplus s_{3}\oplus s_{5}\oplus s_{4}=a_{1}\oplus a_{5}\oplus a_{0}\oplus a_{2}\oplus a_{7}\oplus a_{2}\oplus a_{1}\oplus a_{3}\oplus a_{6}\oplus a_{4}\oplus a_{6}\oplus a_{1}\oplus a_{7}\oplus a_{1}\oplus a_{6}\oplus a_{0}=a_{5}\oplus a_{3}\oplus a_{6}\oplus a_{4}$\\
			\hline	
			\tiny  $ k_{3}$ &\tiny $s_{2}\oplus s_{4}=a_{1}\oplus a_{5}\oplus a_{0}\oplus a_{2}\oplus a_{7}\oplus a_{1}\oplus a_{6}\oplus a_{0}=a_{5}\oplus a_{2}\oplus a_{7}\oplus a_{6}$\\
			\hline	
			\tiny  $ k_{4}$ &\tiny $s_{4}\oplus s_{1}\oplus s_{2}\oplus s_{6}\oplus s_{3}=a_{1}\oplus a_{6}\oplus a_{0}\oplus a_{2}\oplus a_{3}\oplus a_{5}\oplus a_{1}\oplus a_{5}\oplus a_{0}\oplus a_{2}\oplus a_{7}\oplus a_{1}\oplus a_{5}\oplus a_{3}\oplus a_{4}\oplus a_{6}\oplus a_{2}\oplus a_{1}\oplus a_{3}\oplus a_{6}\oplus a_{4}=a_{7}\oplus a_{5}\oplus a_{2}\oplus a_{3}\oplus a_{6}$\\
			\hline	
			\tiny  $ k_{5}$ &\tiny $s_{7}\oplus s_{1}\oplus s_{0}\oplus s_{4}=a_{2}\oplus a_{6}\oplus a_{5}\oplus a_{7}\oplus a_{0}\oplus a_{2}\oplus a_{3}\oplus a_{5}\oplus a_{6}\oplus a_{1}\oplus a_{7}\oplus a_{1}\oplus a_{5}\oplus a_{0}\oplus a_{2}\oplus a_{7}\oplus a_{4}\oplus a_{2}\oplus a_{5}=a_{3}\oplus a_{7}\oplus a_{4}$\\
			\hline	
			\tiny  $ k_{6}$ &\tiny $s_{7}\oplus s_{1}\oplus s_{0}\oplus s_{4}=a_{2}\oplus a_{6}\oplus a_{5}\oplus a_{7}\oplus a_{0}\oplus a_{2}\oplus a_{3}\oplus a_{5}\oplus a_{4}\oplus a_{2}\oplus a_{5}\oplus a_{1}\oplus a_{6}\oplus a_{0}=a_{7}\oplus a_{3}\oplus a_{4}\oplus a_{2}\oplus a_{5}\oplus a_{1}$\\
			\hline	
			\tiny  $ k_{7}$ &\tiny $s_{6}\oplus s_{5}\oplus s_{7}\oplus s_{0}=a_{1}\oplus a_{5}\oplus a_{3}\oplus a_{4}\oplus a_{6}\oplus a_{6}\oplus a_{1}\oplus a_{7}\oplus a_{2}\oplus a_{6}\oplus a_{7}\oplus a_{0}\oplus a_{4}\oplus a_{2}\oplus a_{5}=a_{3}\oplus a_{6}\oplus a_{0}\oplus a_{5}$\\
			\hline		
		\end{tabular}\\
		~\\
	}	
	\vspace{-3mm}
\end{table}

Two efficient matrices $M_{NT}$ and $M_{TN}$ are used for converting elements between normal basis and tower field representations \cite{1}. These conversion matrices lead to small critical path delay and area consumed \cite{1}. The conversion matrices of $M_{NT}$ and $M_{TN}$ are shown below:\vspace{1mm}

\setcounter{MaxMatrixCols}{16}
{\scriptsize \begin{equation*}
	M_{NT}=
	\begin{bmatrix}
	1 & 1 & 0 & 1 & 0 & 0 & 0 & 1 & 1 & 0 & 1 & 1 & 0 & 0 & 0 & 0\\
	0 & 0 & 0 & 0 & 0 & 0 & 1 & 1 & 0 & 1 & 0 & 1 & 0 & 0 & 1 & 0\\
	1 & 0 & 0 & 0 & 1 & 0 & 0 & 1 & 0 & 1 & 0 & 0 & 1 & 0 & 0 & 0\\
	0 & 1 & 1 & 1 & 0 & 0 & 0 & 1 & 0 & 0 & 0 & 0 & 0 & 1 & 1 & 1\\
	1 & 0 & 0 & 1 & 0 & 1 & 0 & 0 & 1 & 0 & 0 & 0 & 0 & 0 & 1 & 0\\
	0 & 0 & 0 & 0 & 0 & 0 & 1 & 0 & 0 & 0 & 0 & 1 & 0 & 1 & 1 & 1\\
	0 & 0 & 1 & 0 & 0 & 0 & 1 & 0 & 0 & 0 & 1 & 1 & 1 & 1 & 1 & 0\\
	0 & 0 & 1 & 0 & 0 & 0 & 1 & 0 & 0 & 1 & 0 & 0 & 0 & 1 & 0 & 1\\
	1 & 0 & 1 & 1 & 0 & 0 & 0 & 0 & 1 & 1 & 0 & 1 & 0 & 0 & 0 & 1\\
	0 & 1 & 0 & 1 & 0 & 0 & 1 & 0 & 0 & 0 & 0 & 0 & 0 & 0 & 1 & 1\\
	0 & 1 & 0 & 0 & 1 & 0 & 0 & 0 & 1 & 0 & 0 & 0 & 1 & 0 & 0 & 1\\
	0 & 0 & 0 & 0 & 0 & 1 & 1 & 1 & 0 & 1 & 1 & 1 & 0 & 0 & 0 & 1\\
	1 & 0 & 0 & 0 & 0 & 0 & 1 & 0 & 1 & 0 & 0 & 1 & 0 & 1 & 0 & 0\\
	0 & 0 & 0 & 1 & 0 & 1 & 1 & 1 & 0 & 0 & 0 & 0 & 0 & 0 & 1 & 0\\
	0 & 0 & 1 & 1 & 1 & 1 & 1 & 0 & 0 & 0 & 1 & 0 & 0 & 0 & 1 & 0\\
	0 & 1 & 0 & 0 & 0 & 1 & 0 & 1 & 0 & 0 & 1 & 0 & 0 & 0 & 1 & 0
	\end{bmatrix} 
	\label{q5}
	\end{equation*}}
\setcounter{MaxMatrixCols}{16}
{\scriptsize \begin{equation*}
	M_{TN}=
	\begin{bmatrix}
	1 & 1 & 1 & 0 & 0 & 0 & 1 & 0 & 0 & 0 & 0 & 0 & 1 & 0 & 1 & 1\\
	1 & 1 & 0 & 1 & 1 & 0 & 0 & 0 & 1 & 0 & 0 & 0 & 1 & 0 & 0 & 1\\
	1 & 0 & 1 & 0 & 0 & 1 & 0 & 0 & 1 & 1 & 1 & 0 & 1 & 1 & 0 & 1\\
	0 & 0 & 0 & 0 & 0 & 0 & 0 & 1 & 1 & 0 & 0 & 0 & 1 & 0 & 0 & 0\\
	1 & 0 & 0 & 1 & 1 & 1 & 0 & 0 & 0 & 0 & 0 & 0 & 0 & 0 & 1 & 0\\
	0 & 1 & 0 & 0 & 0 & 0 & 0 & 0 & 0 & 1 & 0 & 1 & 0 & 1 & 0 & 1\\
	0 & 0 & 1 & 1 & 0 & 1 & 0 & 0 & 1 & 0 & 1 & 0 & 0 & 0 & 0 & 0\\
	1 & 1 & 0 & 1 & 0 & 1 & 0 & 1 & 0 & 1 & 0 & 0 & 1 & 1 & 0 & 1\\
	0 & 0 & 0 & 0 & 1 & 0 & 1 & 1 & 1 & 1 & 1 & 0 & 0 & 0 & 1 & 0\\
	1 & 0 & 0 & 0 & 1 & 0 & 0 & 1 & 1 & 1 & 0 & 1 & 1 & 0 & 0 & 0\\
	1 & 1 & 1 & 0 & 1 & 1 & 0 & 1 & 1 & 0 & 1 & 0 & 0 & 1 & 0 & 0\\
	1 & 0 & 0 & 0 & 1 & 0 & 0 & 0 & 0 & 0 & 0 & 0 & 0 & 0 & 0 & 1\\
	0 & 0 & 0 & 0 & 0 & 0 & 1 & 0 & 1 & 0 & 0 & 1 & 1 & 1 & 0 & 0\\
	0 & 1 & 0 & 1 & 0 & 1 & 0 & 1 & 0 & 1 & 0 & 0 & 0 & 0 & 0 & 0\\
	1 & 0 & 1 & 0 & 0 & 0 & 0 & 0 & 0 & 0 & 1 & 1 & 0 & 1 & 0 & 0\\
	0 & 1 & 0 & 0 & 1 & 1 & 0 & 1 & 1 & 1 & 0 & 1 & 0 & 1 & 0 & 1
	\end{bmatrix} 
	\label{q6}
	\end{equation*}}

The critical path delay and the number of XOR gates for the proposed implementation of $M_{NT}$ and $M_{TN}$ using the optimizing of equations are (3$T_X$ and 57) and (3$T_X$ and 60), respectively, where $T_X$ denotes the time delay of a 2-input XOR gate. On the other hand, the results (critical path delay and the number of XOR gates) for implementation of $M_{NT}$ and $M_{TN}$ matrices in the work \cite{1} are equal to (3$T_X$ and 76) and (4$T_X$ and 84), respectively.

\section{Proposed structure of 16-bit S-box}
\label{PS}

The proposed structure of 16-bit S-box is constructed by an inversion operation and an affine transformation. The inversion-based 8-bit S-box structures with affine transformations and isomorphic mapping are designed in many ISO/IEC standard ciphers, such as AES, Camellia, and CLEFIA \cite{BRI2}. An S-box with not efficient structure leads to a large hardware consumption in a cryptographic application. Therefore, the design and implementation of the S-boxes based on a lightweight and high-security structure is an important subject in the cryptosystems. In this paper, the composite field $\F_{(((2^2)^2)^2)^2}$ is utilized for the implementation of field inversion. The implementation of the proposed S-box using a composite field reduces the hardware resources required for S-box computation. To minimize the cost of the S-box operation, the following steps are employed:\vspace{1mm}

\textbf{Step 1:} Compute the field inversion over the composite field $\F_{(((2^2)^2)^2)^2}$.\vspace{1mm}

\textbf{Step 2:} Apply an area-efficient affine transformation to the output of the inversion.\vspace{1mm}

\begin{figure}[!t]
	\centering
	\includegraphics[scale=0.3]{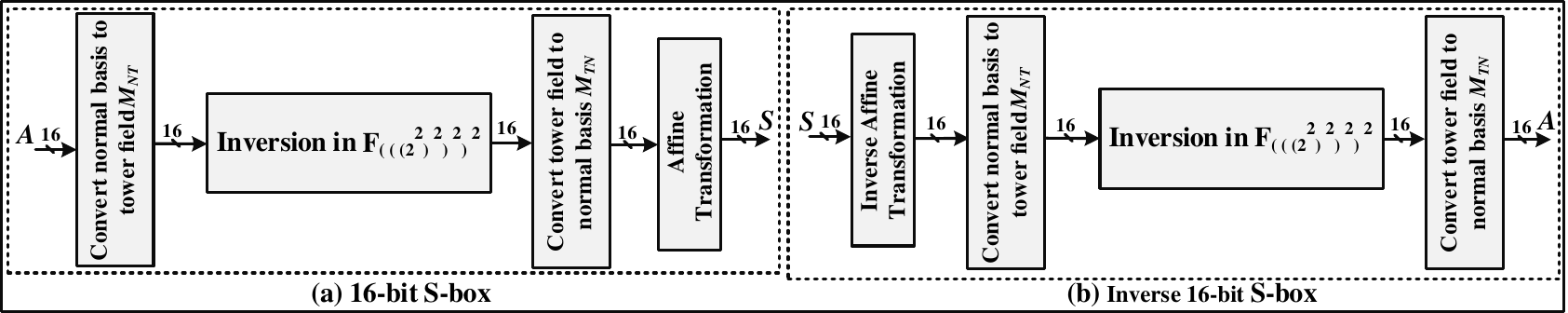}\vspace{-2mm}
	\caption{The proposed structure of 16-bit S-box (a) and S-box$^{-1}$ (b).}
	\label{fig:F6}
	\vspace{-6mm}
\end{figure}

The proposed structure of the 16-bit S-box and its inverse (S-box$^{-1}$) is shown in Figs. \ref{fig:F6} (a) and (b), respectively. In the S-box operation, the 16-bit input is first converted from the normal basis to the tower field representation. Then, the inversion operation over $\mathbb{F}_{(((2^2)^2)^2)^2}$ is performed. Next, the result is converted back from the tower field representation to the normal basis, and finally, the affine transformation $AT(A) = B = M \times A$ is computed, where $A = (a_{15}, a_{14}, \dots, a_1, a_0)$, $B = (b_{15}, b_{14}, \dots, b_1, b_0)$, and $M$ is a $16 \times 16$ matrix. Similarly, in the inverse of the S-box (S-box$^{-1}$), the inversion (with conversion between the normal basis and the tower field representations) is performed after the inverse affine transformation $AT^{-1}(B) = N \times A$, where $N$ is a $16 \times 16$ matrix. Two key factors significantly influence the complexity of the S-box: the hardware architecture of the field elements and the affine transformation. Consequently, it is possible to design an affine transformation that facilitates an efficient implementation. In the following, our matrices $M$ and $N$ for the affine and inverse affine transformations are presented:\vspace{1mm}

{\scriptsize \begin{equation*}
	\setcounter{MaxMatrixCols}{16}
	M=
	\begin{bmatrix}
	0 & 0 & 0 & 0 & 0 & 0 & 0 & 0 & 1 & 0 & 1 & 0 & 1 & 0 & 0 & 0\\
	0 & 0 & 0 & 0 & 0 & 0 & 0 & 0 & 0 & 1 & 0 & 1 & 0 & 1 & 0 & 0\\
	0 & 0 & 0 & 0 & 0 & 0 & 0 & 0 & 0 & 0 & 1 & 0 & 1 & 0 & 1 & 0\\
	0 & 0 & 0 & 0 & 0 & 0 & 0 & 0 & 0 & 0 & 0 & 1 & 0 & 1 & 0 & 1\\
	0 & 0 & 0 & 0 & 0 & 0 & 0 & 0 & 1 & 0 & 0 & 0 & 1 & 0 & 1 & 0\\
	0 & 0 & 0 & 0 & 0 & 0 & 0 & 0 & 0 & 1 & 0 & 0 & 0 & 1 & 0 & 1\\
	0 & 0 & 0 & 0 & 0 & 0 & 0 & 0 & 1 & 0 & 1 & 0 & 0 & 0 & 1 & 0\\
	0 & 0 & 0 & 0 & 0 & 0 & 0 & 0 & 0 & 1 & 0 & 1 & 0 & 0 & 0 & 1\\
	1 & 0 & 1 & 0 & 1 & 0 & 0 & 0 & 0 & 0 & 0 & 0 & 0 & 0 & 0 & 0\\
	0 & 1 & 0 & 1 & 0 & 1 & 0 & 0 & 0 & 0 & 0 & 0 & 0 & 0 & 0 & 0\\
	0 & 0 & 1 & 0 & 1 & 0 & 1 & 0 & 0 & 0 & 0 & 0 & 0 & 0 & 0 & 0\\
	0 & 0 & 0 & 1 & 0 & 1 & 0 & 1 & 0 & 0 & 0 & 0 & 0 & 0 & 0 & 0\\
	1 & 0 & 0 & 0 & 1 & 0 & 1 & 0 & 0 & 0 & 0 & 0 & 0 & 0 & 0 & 0\\
	0 & 1 & 0 & 0 & 0 & 1 & 0 & 1 & 0 & 0 & 0 & 0 & 0 & 0 & 0 & 0\\
	1 & 0 & 1 & 0 & 0 & 0 & 1 & 0 & 0 & 0 & 0 & 0 & 0 & 0 & 0 & 0\\
	0 & 1 & 0 & 1 & 0 & 0 & 0 & 1 & 0 & 0 & 0 & 0 & 0 & 0 & 0 & 0
	\end{bmatrix}
	\label{q55}
	\end{equation*}}
{\footnotesize \begin{equation*}
	N=
	\setcounter{MaxMatrixCols}{16}
	\begin{bmatrix}
	0 & 0 & 0 & 0 & 0 & 0 & 0 & 0 & 1 & 0 & 0 & 0 & 1 & 0 & 1 & 0\\
	0 & 0 & 0 & 0 & 0 & 0 & 0 & 0 & 0 & 1 & 0 & 0 & 0 & 1 & 0 & 1\\
	0 & 0 & 0 & 0 & 0 & 0 & 0 & 0 & 1 & 0 & 1 & 0 & 0 & 0 & 1 & 0\\
	0 & 0 & 0 & 0 & 0 & 0 & 0 & 0 & 0 & 1 & 0 & 1 & 0 & 0 & 0 & 1\\
	0 & 0 & 0 & 0 & 0 & 0 & 0 & 0 & 1 & 0 & 1 & 0 & 1 & 0 & 0 & 0\\
	0 & 0 & 0 & 0 & 0 & 0 & 0 & 0 & 0 & 1 & 0 & 1 & 0 & 1 & 0 & 0\\
	0 & 0 & 0 & 0 & 0 & 0 & 0 & 0 & 0 & 0 & 1 & 0 & 1 & 0 & 1 & 0\\
	0 & 0 & 0 & 0 & 0 & 0 & 0 & 0 & 0 & 0 & 0 & 1 & 0 & 1 & 0 & 1\\
	1 & 0 & 0 & 0 & 1 & 0 & 1 & 0 & 0 & 0 & 0 & 0 & 0 & 0 & 0 & 0\\
	0 & 1 & 0 & 0 & 0 & 1 & 0 & 1 & 0 & 0 & 0 & 0 & 0 & 0 & 0 & 0\\
	1 & 0 & 1 & 0 & 0 & 0 & 1 & 0 & 0 & 0 & 0 & 0 & 0 & 0 & 0 & 0\\
	0 & 1 & 0 & 1 & 0 & 0 & 0 & 1 & 0 & 0 & 0 & 0 & 0 & 0 & 0 & 0\\
	1 & 0 & 1 & 0 & 1 & 0 & 0 & 0 & 0 & 0 & 0 & 0 & 0 & 0 & 0 & 0\\
	0 & 1 & 0 & 1 & 0 & 1 & 0 & 0 & 0 & 0 & 0 & 0 & 0 & 0 & 0 & 0\\
	0 & 0 & 1 & 0 & 1 & 0 & 1 & 0 & 0 & 0 & 0 & 0 & 0 & 0 & 0 & 0\\
	0 & 0 & 0 & 1 & 0 & 1 & 0 & 1 & 0 & 0 & 0 & 0 & 0 & 0 & 0 & 0
	\end{bmatrix}
	\end{equation*}}

The proposed formula affine $B=(b_{15},b_{14},...,b_1,b_0)=AT(A)$ and inverse affine $A=(a_{15},a_{14},...,a_1,a_0)=AT^{-1}(B)$ transformations are as follows:\vspace{0.2cm}

$b_{15}=(a_7\oplus a_5\oplus a_3)$, $b_{14}=a_6\oplus a_4\oplus a_2$, $b_{13}=a_5\oplus a_3\oplus a_1$, $b_{12}=a_4\oplus a_2\oplus a_0$, $b_{11}=a_7\oplus a_3\oplus a_1$, $b_{10}=a_6\oplus a_2\oplus a_0$, $b_{9}=a_7\oplus a_5\oplus a_1$, $b_{8}=a_6\oplus a_4\oplus a_0$, $b_{7}=a_{15}\oplus a_{13}\oplus a_{11}$, $b_{6}=a_{14}\oplus a_{12}\oplus a_{10}$, $b_{5}=a_{13}\oplus a_{11}\oplus a_{9}$, $b_{4}=a_{12}\oplus a_{10}\oplus a_{8}$, $b_{3}=a_{15}\oplus a_{11}\oplus a_{9}$, $b_{2}=a_{14}\oplus a_{10}\oplus a_{8}$,$b_{1}=a_{15}\oplus a_{13}\oplus a_{9}$, $b_{0}=a_{14}\oplus a_{13}\oplus a_{8}$.\vspace{0.2cm}

$a_{15}=b_7\oplus b_3\oplus b_1$, $a_{14}=b_6\oplus b_2\oplus b_0$, $a_{13}=b_7\oplus b_5\oplus b_1$, $a_{12}=b_6\oplus b_4\oplus b_0$, $a_{11}=b_7\oplus b_5\oplus b_3$, $a_{10}=b_6\oplus b_4\oplus b_2$, $ a_{9}=b_5\oplus b_3\oplus b_1$, $a_{8}=b_4\oplus b_2\oplus b_0$, $a_{7}=b_{15}\oplus b_{11}\oplus b_{9}$, $a_{6}=b_{14}\oplus b_{10}\oplus b_{8}$, $a_{5}=b_{15}\oplus b_{13}\oplus b_{9}$, $a_{4}=b_{14}\oplus b_{12}\oplus b_{8}$, $a_{3}=b_{15}\oplus b_{13}\oplus b_{11}$, $a_{2}=b_{14}\oplus b_{12}\oplus b_{10}$, $a_{1}=b_{13}\oplus b_{11}\oplus b_{9}$, $a_{0}=b_{12}\oplus b_{10}\oplus b_{8}$.\vspace{0.2cm} 

\begin{figure}[!t]
	\centering
	\includegraphics[scale=0.4]{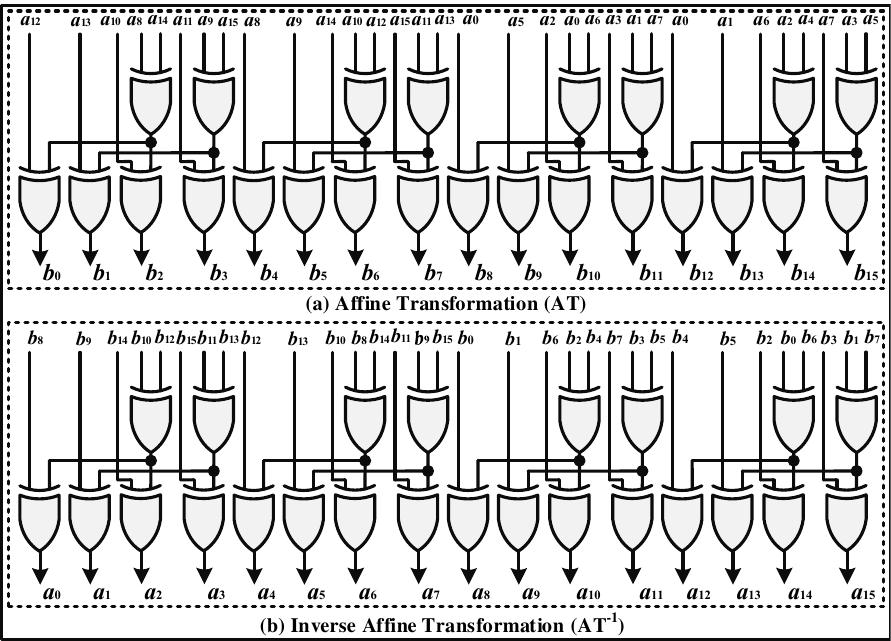}\vspace{-2mm}
	\caption{The structures of proposed affine transformation $AT(A)$ (a) and  inverse affine transformation $AT^{-1}(B)$ (b).}
	\label{fig:F11}
	\vspace{-6mm}
\end{figure}

Fig. \ref{fig:F11} (a) and (b) show the circuits of proposed affine transformation $AT(A)$ and inverse affine transformation $AT^{-1}(B)$, respectively. The $AT(A)$ and $AT^{-1}(B)$ transformations are implemented by 24 2-input XOR gates with the critical path delay of $2T_X$, where $T_X$ is the time delay of a 2-input XOR gate.

\section{Security Results Comparison}
\label{S}

In this section, some of the properties and concepts and required definitions to study the rest of the paper are provided.
\subsection{Math preliminareis} \label{subsec-math-found}
Suppose $m,n$ be two positive integers and $p$ be a prime number. A Galois field also known as a finite set of elements with two 
operations of addition and scalar multiplication with special properties. 
On any both binary vectors $\textbf{x}=(x_{1}, \cdots,x_{n})$ and $\textbf{y}=(y_{1}, \cdots,y_{n})$ in $\mathbb{F}_{2}^{n}=GF(2^{n})$, we define binary addition with $\textbf{x} \oplus \textbf{y}=(x_{1} \oplus y_{1}, \cdots, x_{n} \oplus y_{n} )$ and dot product with $<\textbf{x} \ldotp \textbf{y}>=x_{1}y_{1}+ \cdots +x_{n}y_{n}$. 
\begin{definition} 
	The number of ones in $\textbf{x} \in \mathbb{F}_{2}^{n}$ is called the Hamming weight of $\textbf{x}$ and is represented by $Hw(\textbf{x})$. Also for each  $\textbf{x} \in \mathbb{F}_{2}^{n}$, the set $supp(\textbf{x})$ contains the position of ones in $\textbf{x}$.
\end{definition}
\begin{definition}\label{bf:def} 
	A Boolean function $f$ of $n$ variables is a function of the form $f:\mathbb{F}_{2}^{n} \mapsto  \mathbb{F}_{2}$ (i.e. $F \in \mathcal{B}_{n}$) . Its value vector is the binary vector $v_{f}$ of length $2^{n}$ composed of all $f (\textbf{x})$ when $\textbf{x} \in \mathbb{F}_{2}^{n}$.
\end{definition}
\begin{definition} 
	For an $n$-variable Boolean function $f:\mathbb{F}_{2}^{n} \mapsto  \mathbb{F}_{2}$, the Truth Table (TT) consists of $2^{n}$ rows, used to represent the output of $f$ for all possible input $\mathbf{x}$. 
	So it can be represented in the form of a binary column vector with dimension  $2^{n}$ as form \ref{BF-TT},
	\begin{equation} \label{BF-TT}
	\begin{pmatrix}
	f(0) \\
	\vdots \\
	f(2^{n}-1) \\
	\end{pmatrix}
	\end{equation}
\end{definition}
\begin{definition}
	The number of ones in the TT of a Boolean function $f$ is called the Hamming weight of $f$ and is represented by $Hw(f)$. Obviousely For each Boolean function, $Hw(f)=|supp(f)|$.
\end{definition}

\begin{definition}\label{bf:ANF}
	For any Boolean function $f$, one of the most commonly representations is the multivariate polynomial representation also called the Algebraic Normal Form (ANF) of $f$ that is,
	\begin{equation} 
	f(\textbf{x})=f(x_{1},x_{2},\ldots,x_{n})\!=\!\bigoplus _{I\subseteq \{1,\ldots,n\}}c_{I}\left({\prod _{i\in I}x_{i}}\right),\hspace {0.2cm} c_{I}\in \mathbb {F}_{2}.
	\end{equation}
	
	where the sums include all monomials formed by the variables \cite{van2014encyclopedia}.
\end{definition}
\begin{definition}
	The derivative function of an $n$-variable Boolean function $f$ respect to $\textbf{u} \in \mathbb{F}_{2}^{n}$, is defined as $D_{\textbf{u}}f=f(\textbf{x}) \oplus f(\textbf{x}+ \textbf{u})$ \cite{van2014encyclopedia}.
\end{definition}
Walsh-Hadamard Transform (WHT)\footnote{some authors also call it Walsh transform or Hadamard transform} is a powerful tool for analyzing Boolean functions. It transforms a Boolean function into a different domain where its properties can be analyzed more easily. 
\begin{definition}\label{bf:WHT} 
	A Boolean function $f$ is uniquely determined by its Walsh transform, which is an integer-valued function that can be defined for all $\mathbf{u} \in \mathbb{F}_{2}^{n}$
	\begin{equation}
	W_{f}(\textbf{u}) = \sum_{\textbf{x} \in \mathbb{F}_{2}^{n}} (-1)^{f(\textbf{x}) \oplus <\textbf{u} , \textbf{x}>}
	\end{equation}
\end{definition}
\begin{definition} 
	A Vectorial Boolean Function (VBF) is an extended form of a Boolean function and is a function of the form $F:\mathbb{F}_{2}^{n} \mapsto \mathbb{F}_{2}^{m}$ ( for $m>1$) which maps $n$ binary input vectors to $m$ binary output vectors and sometimes called an $(n,m)$-function (i.e. $F \in \mathcal{B}_{n}^{m}$).
\end{definition}
\begin{definition}\label{vbf:S-def} 
	Any S-box $S \in \mathcal{B}_{n}^{m}$ is a VBF. 
\end{definition}
For any VBF, each output bit can be seen as a Boolean function of the input bits, so $S$ can be defined at every $\textbf{x} \in \mathbb{F}_{2}^{n}$ by $S(\textbf{x})=(f_{1}(\textbf{x}),f_{2}(\textbf{x}),\cdots,f_{m}(\textbf{x}))$. In this case, each $f_{j}(\textbf{x})$ for $j=1, \cdots,m$ is called $j$-th component (or coordinate) function of $S$. 
\begin{definition}
	The TT of $S=(f_{1}, \cdots,f_{m})$ can be shown in the form of a matrix with dimensions $2^{n} \times m$, and its $j$-th column represents the TT of $j$-th component function for $1 \leq j \leq m$. 
	The TT consists of $2^{n}$ rows, used to represent the output of S-box for all possible combinations of its input values. So it can be represented in the form of a matrix as form \ref{S-TT},
	\begin{equation} \label{S-TT}
	\begin{pmatrix}
	f_{1}(v_{0}) \cdots \cdots f_{m}(v_{0}) \\
	\cdots \\
	\cdots  \\
	f_{1}(v_{2^{n}-1}) \cdots  f_{m}(v_{2^{n}-1})\\
	\end{pmatrix}
	\end{equation}
\end{definition}
The ANF of an S-box 
shows which input bits contribute to each output bit through simple operations like XOR. This representation helps in understanding the S-box’s behavior and how it transforms data, which is essential for analyzing its security.
\begin{definition}\label{vbf:ANF} 
	The ANF of an S-box can be expressed as a vector of the $m$ Boolean functions $f_{k}$ as follows:
	\begin{equation} \tiny
	f_{k}(x_{1},\cdots,x_{n})=c_{0}^{(k)}	 \bigoplus_{1 \leq i \leq n} c_{i}^{(k)}x_{i} \bigoplus_{1 \leq i , j \leq n} c_{ij}^{(k)}x_{i}x_{j}\bigoplus_{\cdots} \cdots \bigoplus_{\cdots} c_{12\cdots n}^{(k)}x_{1}x_{2}\cdots x_{n}
	\end{equation}
	where all coefficients belong to $\mathbb{F}_{2}$.
\end{definition}
The WHT of an S-box  quantifies the degree of nonlinearity of the S-box and is defined as a mathematical transformation that measures the correlation between the output of the S-box and all possible linear functions over the binary field. 

\begin{definition}\label{vbf:WHT} 
	The WHT of an S-box $S \in \mathcal{B}_{n}^{m}$ is defined as:
	\begin{equation}
	W_{S}(\textbf{u},\textbf{v}) = \sum_{\textbf{x} \in \mathbb{F}_{2}^{n}} (-1)^{<\textbf{v} \ldotp S(\textbf{x})> \oplus <\textbf{u} \ldotp \textbf{x}>}
	\end{equation}
	where $\textbf{u} \in \mathbb{F}_{2}^{n}$ and $\textbf{v} \in \mathbb{F}_{2}^{m}$ \cite{claude_carlet_2020}.
\end{definition}
\subsection{S-box Design Metrics} \label{subsec-design-prop}
Primary cryptographic properties of Boolean functions and S-boxes represent different security measures against different cryptanalysis attacks. 
Note that these cryptographic properties are not independent of each other, they have some interrelations and some restrictions to each other, and this means that one cannot find a Boolean function or an S-box with all these properties to reach the best extent. 
\subsubsection{Algebraic Degree}
Cryptographic functions must have high algebraic degrees. Indeed, all cryptosystems using Boolean functions for confusion (combining or filtering functions in stream ciphers, functions involved in the S-boxes of block ciphers) can be attacked if the functions have low algebraic degree.
A function with a high algebraic degree is more complex and generally more secure against algebraic attacks. It indicates the function's resistance to being approximated by lower-degree polynomials \cite{carlet2010boolean}. 
\begin{definition}\label{bf:AD}
	Algebraic degree of a Boolean function $f$, is showed by $deg_{f}$ and is defined as the maximum degree of the polynomial in the ANF representation of $f$. 
\end{definition}

\begin{definition}\label{vbf:AD}
	For an S-box $S=(f_{1}, \cdots,f_{m}) \in \mathcal{B}_{n}^{m}$ 
	, the algebraic degree is indicated by $deg_{_{S}}=\max\{deg_{_{f_{j}}} | 1 \leq j \leq m\}$.
\end{definition}
\subsubsection{Differential Uniformity} 
Nyberg introduced the concept of differential uniformity as a measure of an S-box resistance to differential crytanalysis \cite{nyberg1993differentially}.
Differential delta uniformity represents the largest value in the Difference Distribution Table (DDT) without counting the values in the first row. 
Lower differential uniformity values indicate better resistance to differential cryptanalysis, with the theoretical minimum being 2 for $S \in \mathcal{B}_{n}^{n}$ .
\begin{definition}	\label{vbf:du}
	The DDT for any entry pair $(\triangle \textbf{x}, \triangle \textbf{y}) \neq (\textbf{0},\textbf{0})$ where $\triangle \textbf{x}$ is input difference  and $\triangle \textbf{y}$ is output difference define as:
	\begin{equation}
	DDT(\triangle \textbf{x}, \triangle \textbf{y}) = \left| \{\textbf{x} \in \mathbb{F}_{2}^{n} : S(\textbf{x}) \oplus S(\textbf{x} \oplus \triangle \textbf{x}) = \triangle \textbf{y}\} \right|
	\end{equation}
	For an S-box $S \in \mathcal{B}_{n}^{n}$, the differential uniformity $\delta_{S}$ is the maximum value in the DDT for some $\triangle \textbf{y}$, excluding the first row and column:
	\begin{equation}
	\delta_{S} = \max_{\triangle \textbf{x} \neq0, \triangle \textbf{y} \neq 0} DDT(\triangle \textbf{x}, \triangle \textbf{y})
	\end{equation}
	
\end{definition}
\subsubsection{Cycle Structure, Fixed Points and Opposite Fixed Points}
In the context of an S-Box, a fixed point is an input value that, when passed through the S-Box, produces the same output value.
Opposite (also called inverse or negated) fixed points refer to pairs of input values that, when substituted through the S-Box, yield outputs that are complementary or “opposite” to each other in some sense. 
An S-box should have very few, if any fixed or opposite fixed points \cite{mishra2023searching}.

\begin{definition}\label{vbf:FP}
	The cycle structure of an invertible vector Boolean function (permutation) describes the number of cycles and their length. A cycle of $k$ numbers is referred to as a $k$-cycle or a cycle of length $k$ \cite{cubero2015vector}. 	
	
	For an S-box $S \in \mathcal{B}_{n}^{m}$, an input $\textbf{x}$ is a fixed point of the $S$ if $S(\textbf{x})=\textbf{x}$, and an input $\overline{\textbf{x}}$ is an opposite fixed point of the $S$ if $S(\textbf{x}) = \overline{\textbf{x}}$
\end{definition}
The number of fixed points should be minimized to enhance security. An S-box $S \in \mathcal{B}_{n}^{m}$, with no fixed points (i.e.,$FP(S)=0$ ) is ideal, as it prevents trivial attacks. In practice, S-boxes should aim for a low number of fixed points, with a common target being fewer than $2^{n-1}$ fixed points for an S-box $S$.
\subsubsection{Nonlinearity} 
It is a measure of how far a Boolean function $f \in \mathcal{B}_{n}$ is from being an affine Boolean function and defined as the minimum Hamming distance between $f$ and all affine functions \cite{claude_carlet_2020}.  

\begin{definition}\label{bf:NL} 
	The nonlinearity of a Boolean function $f \in \mathcal{B}_{n}$ can be expressed in terms of the WHT coefficients as \cite{claude_carlet_2020}:
	\begin{equation}
	Nl_{f}=2^{n-1}-\dfrac{1}{2}\max_{\textbf{u} \in \mathbb{F}_{2}^{n}} \left| W_{f}(\textbf{u}) \right|
	\end{equation}
\end{definition}

A generalization to VBF of the notion of nonlinearity of Boolean functions has been introduced and studied by Nyberg \cite{nyberg1992construction} and further studied by Chabaud and Vaudenay \cite{chabaud1994links}.
Nonlinearity quantifies how far the S-box deviates from being an affine VBF.
It measures the distance of the S-box to the dependent functions, with higher values indicating better resistance to cryptographic linear analysis \cite{carlet2010boolean,canteaut2016lecture}. 


The nonlinearity of an S-box can be computed using the Walsh-Hadamard Transform (WHT). 
by definition \ref{vbf:WHT}, the minimum Hamming distance of all functions $<\textbf{v} \ldotp S(\textbf{x})>$ with affine functions $<\textbf{u} \ldotp \textbf{x}>$ is called the nonlinear degree of the S-box and is defined as follows, 
\begin{definition}\label{vbf:NL}
	The nonlinearity of $S \in \mathcal{B}_{n}^{m}$ is given by \cite{crama2010boolean},
	\begin{equation}
	NL_{S} = 2^{n-1} - \frac{1}{2} \max_{\textbf{u} \in \mathbb{F}_{2}^{n}} \left| \sum_{\textbf{x} \in \mathbb{F}_{2}^{n}} (-1)^{S(\textbf{x}) \oplus <\textbf{u} , \textbf{x}>}\right| \ , \ (\textbf{u}\neq \textbf{0})
	\end{equation}
	For an S-box $S \in \mathcal{B}_{n}^{n}$, the theoretical lower bound is $0$ and the upper bound for nonlinearity is $2^{n-1}- 2^{\frac{n}{2}}+2$ for $n$ even and $2^{n-1}- 2^{\frac{n-1}{2}}$ for $n$ odd. 
\end{definition}
\subsubsection{Signal to Nois Rate}
In 2004, Guilley presents DPA Signal to Nois Rate (SNR) measure which is, the first property that characterizes resilience of S-boxes to DPA attacks \cite{guilley2004differential}. 
In DPA-resistant S-boxes, the goal is to minimize the correlation between the power consumption and the intermediate values (such as the output of the S-box). The SNR is a metric that helps quantify this correlation. 

The signal in this context refers to the power consumption variations that are directly related to the secret data (e.g., the key or intermediate values). For a DPA attack to be successful, the attacker needs to distinguish these variations from random noise. The noise refers to the power consumption variations that are not related to the secret data. These variations can come from various sources, such as environmental factors, circuit noise, and other non-data-dependent power consumption.




\begin{definition}\label{vbf:SNR}
	For an S-box $S \in \mathcal{B}_{n}^{m}$ with $S=(f_{1},f_{2}, \cdots,f_{m})$, The SNR can be defined as \cite{picek2016evolutionary}: 
	\begin{equation}
	SNR_{DPA}(S)= n 2^{2m} \left(  \sum_{k} \left( \sum_{i=1}^{m} (-1)^{\widehat{f}_{i}(k)}  \right) \right) 
	\end{equation}
	where $\widehat{f}(k)=\sum_{x}(-1)^{<x \ldotp k>}f(x)$.
\end{definition}
\subsubsection{Transparency Order}
The notion of transparency order, proposed by Prouff \cite{prouff2005dpa} and then redefined by Chakraborty et al. \cite{chakraborty2017redefining}, is a property that attempts to characterize the resilience of cryptographic algorithms against DPA attacks. A lower transparency order indicates better resistance, meaning that the function's output does not reveal much information about the input or the key used.
\begin{definition}\label{bf:TO}
	Suppose  $f \in \mathcal{B}_{n}$.
	The transparency order of $f$ denoted by $To_{f}$ is defined as \cite{chakraborty2017redefining}:
	\begin{equation}\label{To_f}
	To_{f}=\max_{\beta \in \mathbb{F}_{2}^{n}} \left(  \left| 1- 2 Hw(\beta) \right| - \dfrac{1}{2^{2n}-2^{n}} 
	\sum_{\alpha \in \mathbb{F}_{2}^{n}} \left|   W_{D_{\textbf{a}}f(\textbf{0},\textbf{v})}	\right| \right)
	\end{equation}
\end{definition}
Transparency order is a relatively new metric that quantifies an S-box's resistance against differential power analysis (DPA) attacks by measuring the predictability of power consumption differences \cite{10.1088/1402-4896/adadab}. In simpler terms, it gauges an S-box’s ability to keep its secrets even when adversaries eavesdrop on its power consumption or electromagnetic radiation. 

\begin{definition}\label{vbf:TO}
	Suppose  $S \in \mathcal{B}_{n}^{m}$.
	The transparency order TO of $S$ denoted by $TO_{S}$ is defined as \cite{chakraborty2017redefining}:
	\begin{equation} \tiny
	TO_{S}=\max_{\beta \in \mathbb{F}_{2}^{n}} \left(  \left| m- 2 Hw(\beta) \right| - \dfrac{1}{2^{2n}-2^{n}} \sum_{\alpha \in \mathbb{F}_{2}^{n}} \left|  \sum_{\textbf{v} \in \mathbb{F}_{2}^{m}, Hw(\textbf{v})=1} (-1)^{<\textbf{v},\beta>} W_{D_{\textbf{a}}S}(\textbf{0},\textbf{v})	\right| \right)
	\end{equation}
	
\end{definition}
Over time, TO got a makeover. Chakraborty et al. analyse the definition from scratch, modify it and finally provide a definition with better insight that can theoretically capture \cite{chakraborty2017redefining}. 
Now it’s the reVisited transparency order (VTO). When designing lightweight cryptographic algorithms (e.g. IoT devices), TO/VTO ensures S-boxes can pirouette gracefully even under SCA scrutiny. 


As seen in Table \ref{T:TT4}, the cryptographic properties for the proposed method are better than or equal to the best S-boxes. 


\begin{table}[!t]
	\centering
	\captionsetup{justification=centering}
	{\scriptsize	
		\caption{Security results analysis for different 16-bit S-boxes.}	\vspace{-0.3cm}
		\centering 
		\label{T:TT4}
		
		\begin{tabular}{|c|c|c|c|c|c|}
			\hline
			\tiny  \bf Works & \tiny   \shortstack{\bf NL} & \tiny   \bf \shortstack{DU}&  \tiny \bf  \shortstack{AD}   & \tiny   \bf \shortstack{TO} & \tiny   \bf \shortstack{SNR}  \\
			\hline
			\tiny  gold(16,1) &\tiny 32512  &\tiny 2  &\tiny 2  &\tiny 15.9998	&\tiny 	3.50e-08	 \\
			\hline			
			\tiny  gold(16,2) &\tiny 32256  &\tiny 4  &\tiny 2 &\tiny 15.9989	&\tiny 	3.33e-08	 \\
			\hline
			\tiny  kasami(16,1)  &\tiny 32512   &\tiny  2  &\tiny  2 &\tiny 15.9998	&\tiny 3.50e-08		 \\
			\hline            			
			\tiny  kasami(16,2)  &\tiny 31232   &\tiny  12  &\tiny  3  &\tiny 15.9787	&\tiny 3.42e-08		 \\
			\hline 
			\tiny  kasami(16,3)  &\tiny 32512   &\tiny  2  &\tiny  4  &\tiny 15.9894	&\tiny 	3.50e-08	 \\
			\hline 	
			\tiny  \cite{r23}  &\tiny 31986   &\tiny  18  &\tiny  15  &\tiny   -  &\tiny   146.423           \\
			\hline 	
			\tiny   This work &\tiny  32512 &\tiny  4 &\tiny  15  &\tiny  15.9875	&\tiny 	0.34e-08	\\
			\hline
			\tiny   Optimal  &\tiny  High &\tiny  Low &\tiny  High &\tiny  Low	&\tiny 	High	\\
			\hline
			
		\end{tabular}\\
		~\\
		NL: Nonlinearity; DU:Differential Uniformity; AD:Algebraic Degree; TO: Transparency Order; SNR: Signal to Noise Ratio. \\ The S-boxes for rows 1 through 5 were generated by SageMath 9.0 and evaluated by Sbox Evaluation Tool (SET) \cite{picek-set}.
	}	
	\vspace{-5mm}
\end{table}

\section{Hardware Results Comparison}
\label{H}

In this section, we compare the hardware complexity of the proposed 16-bit S-box with other 16-bit S-boxes. Our ASIC implementations provide area gate equivalents (GEs), and delay results for the 65 nm CMOS technology using Synopsys Design Compiler. The number of required logic gates and the critical path delay of the proposed field operations and the original operations are compared in Table \ref{T:T5}. For example, the field inversion unit $I_{16}$ in the proposed work and the work in \cite{1} requires (343 XOR gates and 166 NAND/NOR gates) and (427 XOR gates and 117 AND gates), respectively. Additionally, the critical path delay in the proposed field inversion unit is 15 XOR gate delays shorter compared to the work \cite{1}. Based on implementation results, the area consumed and delay of the proposed inversion unit $I_{16}$ and the work \cite{1}, for the 65 nm CMOS technology, are equal to (852 GEs, 3.02 ns) and (1684 GEs, 5.02 ns), respectively. The number of logic gates in the proposed structures is comparable to that in the original circuits and other works.

\begin{table}[!t]
	\centering
	\captionsetup{justification=centering}
	{\scriptsize	
		\caption{Hardware results of the proposed and original field operations.}	\vspace{-0.3cm}
		\centering 
		\label{T:T5}
		
		\begin{tabular}{|c|c|c|c|c|c|}
			\hline
			\tiny  \bf Works & \tiny \bf \shortstack{XOR} & \tiny \bf \shortstack{AND} & \tiny \bf \shortstack{NAND/\\NOR} & \tiny \bf CPD\\
			\hline			
			\tiny  $M_2$ &\tiny  4 &\tiny  3  &\tiny ---  &\tiny $2T_X+T_A$  \\
			\hline                       
			\tiny  Proposed $M_2$ &\tiny  3 &\tiny  ---  &\tiny 4  &\tiny $2T_X+T_{NA}$ \\
			\hline  
			\tiny  $M_4$ &\tiny  21 &\tiny  9  &\tiny ---  &\tiny $5T_X+T_A$  \\
			\hline              
			\tiny  Proposed $M_4$ &\tiny  17 &\tiny ---  &\tiny 12  &\tiny $3T_X+T_{NA}$  \\
			\hline            
			\tiny  $M_8$ &\tiny  107 &\tiny  27  &\tiny ---  &\tiny $12T_X+T_A$  \\
			\hline              
			\tiny  Proposed $M_8$ &\tiny  82 &\tiny  --- &\tiny 36  &\tiny $8T_X+T_{NA}$  \\
			\hline  
			\tiny  $S_4$ &\tiny  7 &\tiny  ---  &\tiny ---  &\tiny $3T_X$  \\
			\hline              
			\tiny  Proposed $S_4$ &\tiny  6 &\tiny --- &\tiny ---  &\tiny $2T_X$  \\
			\hline    
			\tiny  $S_8$ &\tiny  31 &\tiny  ---  &\tiny ---  &\tiny $7T_X$  \\
			\hline              
			\tiny  Proposed $S_8$ &\tiny  15 &\tiny --- &\tiny ---  &\tiny $3T_X$  \\
			\hline 
			\tiny  $I_4$ &\tiny  17 &\tiny  9  &\tiny ---  &\tiny $5T_X+2T_A$  \\
			\hline              
			\tiny  Proposed $I_4$ &\tiny  2 &\tiny --- &\tiny 22  &\tiny $2T_{NA}+3T_{NO}+T_N$  \\
			\hline   
			\tiny  $I_8$ &\tiny  100 &\tiny  36  &\tiny ---  &\tiny $17T_X+3T_A$  \\
			\hline              
			\tiny  Proposed $I_8$ &\tiny  64 &\tiny  --- &\tiny 58  &\tiny $7T_X+4T_{NA}+3T_{NO}+T_N$  \\
			\hline 
			\tiny  $I_{16}$ &\tiny  496 &\tiny  117  &\tiny ---  &\tiny $43T_X+4T_A$  \\
			\hline 
			\tiny \cite{1} $I_{16}$ &\tiny  427 &\tiny  117  &\tiny ---  &\tiny $39T_X+4T_A$  \\
			\hline              
			\tiny  Proposed $I_{16}$ &\tiny  343 &\tiny  --- &\tiny 166  &\tiny $24T_X+6T_{NA}+3T_{NO}+T_N$  \\
			\hline 		                      			
		\end{tabular}\\
		~\\
		 $T_A$, $T_{NA}$, $T_X$, $T_{NO}$, $T_{N}$ denote the time delay of a 2-input AND gate, 2-input NAND gate, 2-input NOR gate, and NOT gate, respectively.
	}	
	\vspace{-6mm}
\end{table}

Hardware results of the proposed 16-bit S-box and other works are presented in Table \ref{T:T6}. The critical path delay of the proposed 16-bit S-box is equal to $32T_X+6T_{NA}+3T_{NO}+T_N$. We implemented the proposed S-box using 180 nm and 65 nm CMOS standard cell libraries. The hardware implementation results of the 16-bit S-boxes presented in the proposed work, as well as the works in \cite{r19} and \cite{r9}, in 65nm technology, are 1135 GEs, 3002 GEs, and 1322 GEs, respectively. In the works \cite{r19} and \cite{r9}, the timing specifications related to the S-boxes were not provided. However, based on the obtained results, the critical path delay in the proposed substitution box is 1.176 ns. The number of GE in the proposed work have a 62.19\% and 14.15\% improvement than that of the works \cite{r19} and \cite{r9}, respectively. The results show that the proposed 16-bit S-box has a low area compared to other S-boxes.

\begin{table}[!t]
	\centering
	\captionsetup{justification=centering}
	{\scriptsize	
		\caption{Hardware results of the proposed 16-bit S-box and other works.}	\vspace{-0.3cm}
		\centering 
		\label{T:T6}
		
		\begin{tabular}{|c|c|c|c|c|c|}
			\hline
			\tiny  \bf Works & \tiny \bf \shortstack{XOR/\\XNOR} & \tiny \bf \shortstack{AND/\\OR} & \tiny \bf \shortstack{NAND/\\NOR} \\
			\hline			
			\tiny \cite{r7} &\tiny  1238 &\tiny  144  &\tiny ---    \\
			\hline  
			\tiny \cite{r9} &\tiny  420 &\tiny  251  &\tiny ---    \\
			\hline   
			\tiny \cite{r19} &\tiny  1238 &\tiny  144  &\tiny ---   \\
			\hline            
			\tiny  This work &\tiny  484 &\tiny --- &\tiny 166   \\
			\hline 		                      			
		\end{tabular}\\
		~\\

	}	
	\vspace{-6mm}
\end{table}

\section{Conclusion}
\label{sec-end}

In this paper, a very compact and high-secure 16-bit S-box is presented. The proposed S-box is based on a field inversion unit as a core element in the composite field $\F_{(((2^2)^2)^2)^2}$ with a low-cost affine transformation. The original field $\F_{2^{16}}$ is converted into the tower fields such as $\F_{(((2^2)^2)^2)^2}$, $\F_{((2^2)^2)^2}$, and $\F_{(2^2)^2}$. The operations in these sub-fields are optimally rewritten to be more compact compared to those in the original field. The sub-blocks of some field operation are merged for generating a unified and low area circuit. A low-cost affine transformation with low hardware and low delay is used in the S-box. The hardware results show that the proposed 16-bit S-box has lower hardware consumption than that of other 16-bit S-boxes. The security level of the proposed S-box is acceptable based on security analysis.



\begin{thebibliography}{8}
	


	\bibitem{r17}
	{Prevost, T., Martin, B.}, {A 10-bit S-box generated by Feistel construction from cellular automata}, {\it Cryptology {ePrint}}, Archive, Paper 2025/457, 2025.
	
	\bibitem{5}
	{Razaq, A., Al-Olayan, H.A., Ullah, A., Riaz, A., and Waheed, A.}, {A Novel Technique for the Construction of Safe Substitution Boxes Based on Cyclic and Symmetric Groups},  {\it Security and Communication Networks}, Vol. 2018, 2018, pp. 1-10.
	
	\bibitem{6}
	{Tian, Y., and Lu, Z.}, {Chaotic S-Box: Intertwining Logistic Map and Bacterial Foraging Optimization}, {\it Mathematical Problems in Engineering}, Vol. 2017, 2017, pp. 1-12.
	

	\bibitem{8}
	{Shuai, L., Wang, L., Miao, L., and Zhou, X.}, {S-Boxes Construction Based on the Cayley Graph of the Symmetric Group for UASNs},  {\it IEEE Access}, Vol. 7, 2019, pp. 38826-38832.
	
	\bibitem{11}
	{Asif Khan, M., Ali, A., Jeoti, V., and Manzoor, S.}, {A Chaos-Based Substitution Box (S-Box) Design with Improved Differential Approximation Probability (DP)},  {\it Iranian Journal of Science and Technology}, Vol. 42, Iss. 2, 2018, pp. 219-238.
	
	\bibitem{13}
	{Isa, H., Jamil, N., and Reza Zaba, M.}, {Construction of Cryptographically Strong S-Boxes Inspired by Bee Waggle Dance},  {\it New Generation Computing}, Vol. 34, Iss. 3, 2016, pp. 221-238. 
	
	\bibitem{14}
	{Rafiq, A., and Khan, M.}, {Construction of new S-boxes based on triangle groups and its applications in copyright protection},  {\it Multimedia Tools and Applications}, Vol. 78, 2019, pp. 15527-15544. 
	
	\bibitem{MK}
	{Muhammad Ali, K., and Khan, M.}, {A new construction of confusion component of block ciphers},  {\it Multimedia Tools and Applications}, Vol. 78, 2019, pp. 32585-32604.	
	
	\bibitem{15}
	{Dey, S., and Ghosh, R.}, {A smart review and two new techniques using 4-bit Boolean functions for cryptanalysis of 4-bit crypto S-boxes},  {\it International Journal of Computers and Applications}, Vol. 2018, 2018, pp. 1-19. 
	
	\bibitem{16}
	{Ahmad, M., Doja, M.N., and Sufyan Beg, M.M.}, {ABC Optimization Based Construction of Strong Substitution-Boxes},  {\it Wireless Personal Communications}, Vol. 101, Iss. 3, 2018, pp. 1715-1729.
	
	\bibitem{22}
	{Zahid, A.H., Arshad, M.J.}, {An Innovative Design of Substitution-Boxes Using Cubic Polynomial Mapping},  {\it Symmetry}, Vol. 11, Iss. 3, 2019, pp. 1-10.
	
  

	\bibitem{Imran}
	{Shahzad, I., Mushtaq, Q., and Razaq, A.}, {Construction of New S-Box Using Action of Quotient of the Modular Group for Multimedia Security},  {\it Security and Communication Networks}, Vol. 2019, 2019, pp. 1-10.
	
	\bibitem{Dragan}
	{Lambic, D.}, {S-box design method based on improved onedimensional discrete chaotic map},  {\it Journal of Information and Telecommunication}, Vol. 2, Iss. 2, 2018, pp. 181-191.



	\bibitem{r24}
	{Thakor, V.A., A. Razzaque, M., D. Darji, A., R. Patel, A.,}, {A novel 5-bit S-box design for lightweight cryptography algorithms}, {\it Journal of Information Security and Applications}, Vol. 73, 2023, pp. 1-11.	
	
	\bibitem{BRI}
	{Rashidi, B.}, {Compact and Efficient structure of 8-bit S-box for lightweight cryptography}, {\it Integration, the VLSI Journal}, Vol. 76, 2020, pp. 172-182.
	
	\bibitem{BRI2}
	{Rashidi, B.}, {Lightweight 8-bit S-box and combined S-box/S-box$^{-1}$ for cryptographic applications}, {\it Int J Circ Theor Appl.}, Vol. 49, Iss. 8, 2021, pp. 2348-2362.

	
	\bibitem{r2}
	{Kim, H., et al.}, {A New Method for Designing Lightweight S-Boxes With High Differential and Linear Branch Numbers, and Its Application},  {\it IEEE Access}, Vol. 9, 2021, pp. 150592-150607.

	\bibitem{r4}
	{Stafford, D.F.}, {Evaluating Performance and Efficiency of a 16-bit Substitution Box on an FPGA}, Thesis,  Master of Science, Rochester Institute of Technology, 2021.

	\bibitem{r5}
	{Blocklove, J., Farris, S., Kurdziel, M., Lukowiak, M., and Radziszowski, S.}, {Hardware Obfuscation of the 16-bit S-box in the MK-3 Cipher}, {\it in Proc. the 28$^{th}$ International Conference Mixed Design of Integrated Circuits and Systems}, 2021, Lodz, Poland, pp. 104-109.

	\bibitem{r6}
	{A. Wood , S.}, {Large substitution boxes with efficient combinational}, Thesis,  Master of Science, Rochester Institute of Technology, 2013.
		
	\bibitem{r7}
	{Werner, G., et al.}, {Implementing Authenticated Encryption Algorithm MK-3 on FPGA}, {\it in Proc. IEEE Military Communications Conference}, 2016, Baltimore, MD, USA, pp. 1-6.

	\bibitem{r9}
	{Li, Y., Zhang, W., Lin, Y., Zou, J., Liu, J.}, {A circuit area optimization of MK-3 S-box},  {\it Cybersecurity}, Vol. 7, 2024, pp. 1-11.

	\bibitem{r11}
	{Zidaric, N., Aagaard, M., Gong, G.}, {Hardware optimizations and analysis for the WG-16 cipher with tower field arithmetic},  {\it IEEE Transactions on Computers}, Vol. 68, Iss. 1, 2018, pp. 67-82.

	\bibitem{r12}
	{Zidaric, N.}, {Hardware Implementations of the WG-16 Stream Cipher with Composite Field Arithmetic}, Thesis,  Master of Science, University of Waterloo, 2014.

	\bibitem{r19}
	{A. Wood , S., Radziszowski, S., and Lukowiak, M.}, {Constructing Large S-boxes with Area Minimized Implementations}, {\it in Proc. IEEE Military Communications Conference}, 2015, Tampa, FL, USA, pp. 49-54.

	\bibitem{r23}
	{Xiao-Nian, W., Dao-Rao, D., Yong-Zhuang, W., Run-Lian, Z., Ling-Chen, L.,}, {A 16-bit S-box Design Method Based on Feistel-NFSR Structure},  {\it Journal of Cryptologic Research}, Vol. 10, Iss. 1, 2022, pp. 146-154.


	\bibitem{1}
	{Fan, X., Zidaric, N., Aagaard, M., and Gong, G.}, {Efficient Hardware Implementation of the Stream Cipher WG-16 with Composite Field Arithmetic}, {\it in Proc. the 3rd international workshop on Trustworthy embedded devices}, 2013,  Berlin, Germany, pp. 21-34.
	

		\bibitem{30}
	{Reyhani-Masoleh, A., Taha, M., Ashmawy, D.}, {Smashing the Implementation Records of AES S-box}, {\it  IACR Transactions on Cryptographic Hardware and Embedded Systems}, Vol. 2018, No. 2, 2018, pp. 298-336.
	
	
	
	
	\bibitem{12H}
	{Mozaffari-Kermani, M., and Reyhani-Masoleh, A.}, {A Low-Cost S-box for the Advanced Encryption Standard Using Normal Basis}, {\it in Proc. IEEE International Conference on Electro/Information Technology}, 2009, Windsor, ON, Canada, pp. 52-55.
	
	\bibitem{7_4}
	{Zhang, X., G., and Parhi, K.K.}, {On the optimum constructions of composite field for the AES algorithm},  {\it  IEEE Trans. Circuits Syst. II}, Vol. 53, Iss. 10, 2006, pp. 1153-1157.
	
	\bibitem{41}
	{Jeon, Y.S., Kim, Y.J. and Lee, D.H.}, {A Compact Memory-Free Architecture for the AES Algorithm Using Resource Sharing Methods}, {\it Journal of Circuits, Systems, and Computers}, Vol. 19, No. 5, 2010, pp. 1109-1130.
	
	    \bibitem{carlet2007nonlinearities}
	{Carlet, C. and Ding, C.}, {Nonlinearities of S-boxes}, {\it Finite fields and their applications}, Vol. 13, 2007, pp. 121-135.
	
	\bibitem{van2014encyclopedia}
	{Van T., Henk C. and Jajodia, S.}, {Encyclopedia of cryptography and security}, {\it Springer Science \& Business Media Press}, 2014.
	
	\bibitem{claude_carlet_2020}
	{Carlet C.}, {Boolean Functions for Cryptography and Coding Theory}, 2020.
	
	\bibitem{carlet2010boolean}
	{Carlet C., Crama, Y., Hammer, P.L. },
	{Boolean Functions for Cryptography and Error-Correcting Codes.}, 2010.
	
	\bibitem{nyberg1993differentially}
	{Nyberg, K.}, {Differentially uniform mappings for cryptography}, {\it Springer Press},	1993, pp. 55-64.
	
	\bibitem{mishra2023searching}
	{Mishra, R., Singh, B., Delhibabu, R.}, {Searching for S-boxes with better Diffusion using Evolutionary Algorithm}, {\it Cryptology ePrint Archive}, 2023.
	
	\bibitem{cubero2015vector}
	{Cubero, J. A.}, {Vector Boolean Functions: Applications in Symmetric Cryptography}, {Universidad Polit{\'e}cnica de Madrid, PhD Thesis}, 2015.
	
	\bibitem{canteaut2016lecture}
	{Canteaut, A.}, {Lecture notes on cryptographic Boolean functions}, {\it Inria, Paris, France},	Vol. 3, 2016.
	
	\bibitem{nyberg1992construction}
	{Nyberg, K.}, {On the construction of highly nonlinear permutations}, {\it Workshop on the Theory and Application of of Cryptographic Techniques},	Vol. 3, 1992, pp. 92-98.   
	
	\bibitem{chabaud1994links}
	{Chabaud, F. Vaudenay, S.}, {Links between differential and linear cryptanalysis}, {\it Workshop on the Theory and Application of of Cryptographic Techniques},	Vol. 3, 1994, pp. 356-365.      
	
	\bibitem{crama2010boolean}
	{Crama, Y., Hammer, P.L.}, {Boolean models and methods in mathematics, computer science, and engineering}, {\it Cambridge University Press}, 2010.
	
	\bibitem{guilley2004differential}
	{Guilley, S., Hoogvorst, P., Pacalet, R.}, {Smart Card Research and Advanced Applications VI}, {\it in Proc. IFIP 18th World Computer Congress TC8/WG8. 8 \& TC11/WG11. 2 Sixth International Conference on Smart Card Research and Advanced Applications}, 2004, Toulouse, France, pp. 127-142.
	
	\bibitem{picek2016evolutionary}
	{Picek, S., Carlet, C., Guilley, S., Miller, J., Jakobovic, D.}, {Evolutionary algorithms for boolean functions in diverse domains of cryptography}, {\it Evolutionary computation},	Vol. 24, 2016, pp. 667-694.
	
	\bibitem{picek-set}
	{Picek, S., Batina, L., Jakobović, D., Ege, B., \& Golub, M.}, {S-box, SET, match: a toolbox for S-box analysis}, {\it in Proc. 8th IFIP WG 11.2 International Workshop, Securing the Internet of Things}, Heraklion, Crete, Greece, 2014, pp. 140-149.
	
	\bibitem{prouff2005dpa}
	{Prouff, E.}, {International Workshop on Fast Software Encryption Proceedings, DPA attacks and S-boxes}, {\it Springer Press}, 2005, pp. 424-441.
	
	\bibitem{chakraborty2017redefining}
	{Chakraborty, K., Sarkar, S., Maitra, S., Mazumdar, B., Mukhopadhyay, D., Prouff, E.}, {Redefining the transparency order}, {\it Designs, codes and cryptography}, Vol. 82, 2017, pp. 95-115.
	
	\bibitem{10.1088/1402-4896/adadab}
	{Khadem, B., Gholamzadeh, M., Ahmad, M., Ansari, Z.A.}, {Robust Image Encryption Scheme Based on 6D Hyper-chaos and DPA-resistant S-box}, {\it Physica Scripta}, Vol. 100, 2025.
	
	
\end{thebibliography}
\end{document}